\documentclass[twocolumn]{aastex631}
\usepackage[utf8]{inputenc}
\usepackage{comment}
\usepackage{graphicx,amsmath,gensymb,rotating}
\usepackage{subfigure}
\usepackage{natbib}
\usepackage{xcolor}
\usepackage{threeparttable,booktabs}

\newcommand{\mgii}{\ion{Mg}{2}}
\newcommand{\civ}{\ion{C}{4}}

\newcommand{\sio}{\ion{Si}{4}+\ion{O}{4}] $\lambda1400$}

\newcommand{\ciii}{\ion{C}{3}]}
\newcommand{\ovi}{\ion{O}{6}}
\newcommand{\lya}{Ly$\alpha$}
\newcommand{\hb}{H$\beta$}

\newcommand{\clagn}{EVA~}

\newcommand{\clagno}{EVA}
\newcommand{\ntot}{10,752~}
\newcommand{\nclagn}{100~}
\newcommand{\nclagno}{100}
\newcommand{\nclagnnew}{98~}

\newcommand{\nlit}{1,206~}

\newcommand{\rv}[1]{{#1}}
\newcommand{\rvt}[1]{\textbf{#1}}

\newcommand{\GuoHX}{H.-X.~Guo et al.~\citeyear{GuoH2020}}
\newcommand{\GuoWJ}{W.-J.~Guo et al.~\citeyear{GuoWJ2025a}}

\def\arcsec{$^{\prime\prime}$}
\def\arcmin{$^{\prime}$}

\usepackage{CJKutf8}            % Chinese name

\begin{document}

\title{Emission-line Variable Active Galactic Nuclei at Cosmic Noon from HETDEX}

\author[0000-0001-5561-2010]{Chenxu Liu\begin{CJK*}{UTF8}{gkai} (刘辰旭) \end{CJK*}}
\affiliation{South-Western Institute for Astronomy Research, Key Laboratory of Survey Science of Yunnan Province, Yunnan University, Kunming, Yunnan 650500, People's Republic of China}
\correspondingauthor{Chenxu Liu}
\email{cxliu@ynu.edu.cn}

\author[0009-0009-9343-090X]{Fanchuan Kong\begin{CJK*}{UTF8}{gkai} (孔凡川) \end{CJK*}}
\affiliation{South-Western Institute for Astronomy Research, Key Laboratory of Survey Science of Yunnan Province, Yunnan University, Kunming, Yunnan 650500, People's Republic of China}

% *** Data Acess ***********************************
\author[0000-0002-2307-0146]{Erin Mentuch Cooper}
\affiliation{Department of Astronomy, The University of Texas at Austin, 2515 Speedway, Austin, TX 78712, USA}
\affiliation{McDonald Observatory, The University of Texas at Austin, 2515 Speedway, Austin, TX 78712, USA}

\author[0000-0002-8925-9769]{Dustin Davis}
\affiliation{Department of Astronomy, The University of Texas at Austin, 2515 Speedway, Austin, TX 78712, USA}

% *** Frequent Discussions **************************
\author[0000-0001-9457-0589]{Wei-Jian Guo\begin{CJK*}{UTF8}{gkai} (郭威坚) \end{CJK*}}
\affiliation{Key Laboratory of Optical Astronomy, National Astronomical Observatories, Chinese Academy of Sciences, Beijing 100012, People's Republic of China}

\author[0000-0001-7240-7449]{Donald P. Schneider}
\affiliation{Department of Astronomy \& Astrophysics, The Pennsylvania State University, University Park, PA 16802, USA}
\affiliation{Institute for Gravitation and the Cosmos, The Pennsylvania State University, University Park, PA 16802, USA}

\author[0000-0002-2610-3919]{Liang Xu}
\affiliation{Yunnan Observatories, Chinese Academy of Sciences, Kunming 650011, People's Republic of China}
\affiliation{Key Laboratory for the Structure and Evolution of Celestial Objects, Chinese Academy of Sciences, Kunming 650011, People's Republic of China}

% *** alphabetical sequence **************************
\author[0000-0002-8433-8185]{Karl Gebhardt}
\affiliation{Department of Astronomy, The University of Texas at Austin, 2515 Speedway, Austin, TX 78712, USA}

\author[0000-0001-6717-7685]{Gary J. Hill}
\affiliation{McDonald Observatory, The University of Texas at Austin, 2515 Speedway, Austin, TX 78712, USA}
\affiliation{Department of Astronomy, The University of Texas at Austin, 2515 Speedway, Austin, TX 78712, USA}

\author[0000-0002-0417-1494]{Wolfram Kollatschny}
\affiliation{Institut f\"ur Astrophysik, Universit\"at G\"ottingen, Friedrich-Hund Platz 1, 37077 G\"ottingen, Germany}

\author{Mirko Krumpe}
\affiliation{Leibniz-Institut für Astrophysik Potsdam (AIP), An der Sternwarte 16, 14482 Potsdam, Germany}

\author[0000-0003-3823-8279]{Shiro Mukae}
\affiliation{Department of Astronomy, The University of Texas at Austin, 2515 Speedway, Austin, TX 78712, USA}

\author[0000-0003-2284-8603]{M. C. Powell}
\affiliation{Leibniz-Institut für Astrophysik Potsdam (AIP), An der Sternwarte 16, 14482 Potsdam, Germany}

\author[0000-0003-2575-0652]{Daniel J. Farrow}
\affiliation{E. A. Milne Centre for Astrophysics University of Hull, Cottingham Road, Hull, HU6 7RX, UK}
\affiliation{Centre of Excellence for Data Science,
Artificial Intelligence \& Modelling (DAIM), University of Hull, Cottingham Road, Hull, HU6 7RX, UK}

\shorttitle{Emission-line Variable AGN at $z\sim1.5$}
\shortauthors{Liu et al.}

% ================================================================
\begin{abstract}

We present the first statistical census of emission-line variable active galactic nuclei (\clagno) at cosmic noon by combining untargeted and deep HETDEX spectroscopy with multi-epoch spectra from SDSS, DESI, and LAMOST. Anchoring all candidates to a HETDEX spectroscopic epoch and requiring AGN classification in either the HETDEX or the external epoch(s), we identify a homogeneous sample of \nclagn \clagn at $z\sim1.5$, including \nclagnnew newly identified. Emission-line variability is selected primarily through statistically significant line-flux changes, supplemented by extensive visual inspections using contemporaneous photometric light curves. The resulting incidence fraction is $f_{\rm \clagn}\sim0.9\%$. The rest-frame intervals between spectroscopic epochs span $\sim$1--10 yr, with brightening and dimming events exhibiting statistically indistinguishable characteristic timescales ($\Delta T\sim2.2$ and $\sim2.6$ yr, respectively). A key result is the characterization of the Baldwin effect in the time domain: while many \clagn follow the ensemble Baldwin effect (eBeff) between two epochs, a substantial fraction exhibit apparent anti-eBeff responses. Time-resolved spectroscopy of an individual source reveals that the intrinsic EW--luminosity relation is non-stationary, with the line-to-continuum responsivity systematically evolving from stronger to weaker across successive variability cycles; sparse two-epoch sampling of this evolving intrinsic Baldwin evolution (iBeff) naturally produces both eBeff-like and anti-eBeff behaviors. Finally, \clagn show no strong preference for extreme Eddington ratios but exhibit a mild tendency toward lower $\lambda_{\rm Edd}$ values relative to matched control samples, driven primarily by sources observed in their dim states. Together, these results establish a coherent framework for interpreting emission-line variability in AGN at the peak epoch of cosmic black hole growth.
\end{abstract}

% ====================================================================================
\section{Introduction}\label{sec_intro}

Active galactic nuclei (AGN) are known to exhibit variability across the electromagnetic spectrum, reflecting changes in accretion, obscuration, and the physical state of the circumnuclear environment.
Among the most dramatic manifestations of AGN variability are so-called changing-look active galactic nuclei (CL-AGN), which show pronounced spectral transitions on timescales of months to years, including the appearance or disappearance of broad emission lines.
Such behavior blurs the classical distinction between Type~1 and Type~2 AGN and demonstrates that AGN spectral classifications are not necessarily static.
\rv{Early examples of changing-look behavior were already reported in spectroscopic studies in the 1970s and 1980s \citep[e.g.,][]{Tohline1976, Cohen1986}, well before the modern AGN/quasar distinction became standard.
}

Renewed interest in CL-AGN was sparked by systematic discoveries in large spectroscopic surveys, most notably through multi-epoch Sloan Digital Sky Survey (SDSS; \citealt{York2000}) observations \citep[e.g.,][]{LaMassa2015, MacLeod2016}.
Since then, CL-AGN have been recognized as a rare but powerful probe of AGN variability, \rv{as they are accompanied by large-amplitude changes in both broad emission lines and continuum variability across optical and infrared wavelengths} \citep[e.g.,][]{Sheng2017, Runnoe2016, Graham2020}.
Owing to these dramatic transitions, CL-AGN provide a unique opportunity to study accretion physics in luminous states while also revealing host-galaxy properties when the nuclear emission is diminished.

At low redshifts ($z \lesssim 1$), changing-look transitions are most commonly identified through variability in the Balmer emission lines \citep[e.g.,][]{MacLeod2016, Gezari2017, FrederickS2019, Green2022, Hon2022, Zeltyn2022, Yang2023}.
At higher redshifts, however, the Balmer lines shift out of the optical window, and spectroscopic identification of changing-look behavior relies on ultraviolet emission lines such as \mgii, \ciii, and \civ\ \citep[e.g.,][]{GuoH2019, Ross2020, GuoWJ2024, GuoWJ2025b}.
The behavior of these lines is less well understood.
For instance, \mgii\ has been reported to respond more weakly to continuum variability than the Balmer lines \citep{Sun2015, GuoH2020b, Zeltyn2024}, likely reflecting differences in excitation mechanisms, optical depth, and broad-line region (BLR) geometry, while the responsivity of high-ionization lines such as \civ\ remains debated.
Expanding the sample of high-redshift objects with well-characterized emission-line variability is therefore essential for understanding BLR physics across cosmic time.

Multiple physical mechanisms have been proposed to explain changing-look phenomena.
Rapid changes in the accretion rate can alter the ionizing continuum and drive coordinated variability in broad emission lines \citep[e.g.,][]{Lu2025}, while variable obscuration by clumpy circumnuclear material may also mimic spectral state changes \citep[e.g.,][]{Ricci2023}.
In a small number of cases, transient events such as tidal disruption events (TDEs) have been invoked, most notably for 1ES~1927+654, where the delayed emergence of broad emission lines following a continuum outburst is well explained by a TDE scenario \citep[e.g.,][]{LiR2022}.
Disentangling these mechanisms requires statistical samples that probe both the frequency and timescales of emission-line variability, particularly at higher redshifts where rest-frame timescales are compressed and observational baselines are limited.

\rv{In this work, we adopt the term ``Emission-line Variable AGN'' (\clagno) as an operational designation for AGN that exhibit statistically significant variability in their emission-line properties between spectroscopic epochs.
This category encompasses classical CL-AGN defined by the appearance or disappearance of broad emission lines, as well as closely related objects showing large but incomplete changes in emission-line fluxes or ratios.
The use of \clagn is not intended to redefine existing classes in the literature, where the term CL-AGN remains standard, but rather to provide a unified observational framework suited to systematic sample construction and statistical analysis.}

The Hobby--Eberly Telescope Dark Energy Experiment (HETDEX; \citealt{Gebhardt2021}) provides a unique opportunity to identify \clagn at intermediate to high redshifts.
As an untargeted and deep spectroscopic survey reaching $g\sim25$ \rvt{(AB)}, HETDEX offers uniform spectral coverage independent of source brightness or prior classification.
By pairing HETDEX spectroscopy with multi-epoch spectra from SDSS, DESI, and LAMOST, we construct a large and homogeneous sample of \clagn spanning $0.5 \lesssim z \lesssim 2.5$, with a redshift distribution peaking near cosmic noon ($z\sim1.5$).
This enables the first statistical census of emission-line variability at the epoch when cosmic black hole growth is most active.

This paper is organized as follows.
Section~\ref{sec_identification} describes the construction of the parent sample, the \clagn selection criteria, and the visual inspection procedures.
Section~\ref{sec_result} presents the statistical properties of the \clagn sample, including duty cycle, characteristic timescales, emission-line behavior, brightening-to-dimming ratios, the time-domain Baldwin effect, and Eddington-ratio distributions.
We summarize our findings and discuss their implications for AGN variability and accretion physics in Section~\ref{sec_summary}.

% ====================================================================================
\section{\clagn Identification} \label{sec_identification}

We identify a robust sample of \clagn by combining spectroscopic observations from HETDEX and other large surveys (SDSS, DESI, and LAMOST) (\S\ref{sec_spec}). Starting from a parent sample constructed via complementary \emph{forward} and \emph{reverse} searches that pair external spectroscopic surveys (SDSS, DESI, and LAMOST) with forced HETDEX spectral extractions at the same sky positions, (\S\ref{sec_parent}), we select \clagn candidates mainly based on emission line flux ratios (\S\ref{sec_selection}). 
VIs are conducted to remove candidates affected by poor spectral quality, or flux calibration issues, using both concurrent light curves (\S\ref{sec_vi_phot}) and the spectra (\S\ref{sec_vi_spec}). This results in a final sample of \nclagn \clagn, of which \nclagnnew are newly identified in this work (\S\ref{sec_clagn_sample}).

\subsection{Spectroscopic Data} \label{sec_spec}

\subsubsection{External Spectroscopic Data} \label{sec_external}

\paragraph{SDSS.}
The Sloan Digital Sky Survey (SDSS) is conducted using a 2.5-meter Sloan telescope at Apache Point Observatory and has provided a comprehensive database of quasar and galaxy spectra through its multi-phase projects (SDSS I/II, III, IV, and V; \citealt{York2000, Dawson2013, Abazajian2009, Blanton2017, Kollmeier2026}). 
\rv{Depending on the survey phase,} SDSS spectra typically span a wavelength range of 
$\sim$3600--10400~\AA, with a spectral resolution of $R \sim 2000$ for quasars and galaxies \citep{Smee2013}.
In this study, we utilize the SDSS Data Release 16 Quasar catalog (DR16Q; \citealt{Lyke2020}), which reaches a magnitude limit of $g\sim22.5$, as a primary external spectroscopic epoch for systematically searching \clagn candidates.

\paragraph{DESI.}
\rv{DESI is a Stage~IV ground-based optical spectroscopic survey conducted with the 4\,m Mayall Telescope at Kitt Peak National Observatory \citep{Levi2013,desi2016a,desi2022}. It provides optical spectra obtained with robotic fiber positioning, covering the wavelength range $\sim3600$--$9800$~\AA\ through three
spectrograph arms with spectral resolutions of $R \sim 2000$--$4000$ \citep{desi2016b}.
In this work, DESI DR1 spectra \citep{desi2025} are used as an additional external spectroscopic epoch, particularly for sources lacking repeat coverage in earlier surveys, and to increase the availability of multi-epoch spectroscopy.
}

\paragraph{LAMOST.}
\rv{LAMOST is a 4\,m quasi-meridian reflecting Schmidt telescope designed for wide-field spectroscopic surveys \citep{Cui2012, Zhao2012}. It provides low- to medium-resolution optical spectra ($R \sim 1800$) covering a wavelength range of approximately $3700$--$9000$~\AA\ \citep{Luo2015}. In this work, we incorporate LAMOST spectra compiled by \citet{Jin2023} as supplementary external spectroscopic observations for a subset of sources. These spectra, drawn from multiple LAMOST observing epochs, extend the temporal baseline available for assessing spectroscopic variability and identifying changing-look AGN candidates.
}

% ....................................................
\subsubsection{HETDEX} \label{sec_hetdex}
HETDEX \citep{Gebhardt2021} is an untargeted spectroscopic survey conducted on the 10-meter Hobby--Eberly Telescope\footnote{\url{http://hetdex.org/}} (HET; \citealt{Hill2021}) at McDonald Observatory. \rv{HETDEX is designed to map large-scale structure during the cosmic noon epoch using Ly$\alpha$ emission, targeting the redshift range $1.88 < z < 3.52$, which corresponds to the Ly$\alpha$ line falling within the instrumental wavelength coverage of 3500--5500~\AA.} 
By providing redshifts for approximately one million Lyman-$\alpha$ emitters (LAEs) over $\sim540~\mathrm{deg}^2$, HETDEX aims to measure the cosmic expansion rate to $\sim1\%$ precision. The untargeted spectroscopic observations, with no imaging pre-selection, are carried out using the Visible Integral-field Replicable Unit Spectrograph (VIRUS; \citealt{Hill2021}), which comprises an array of 78 integral field units (IFUs) spanning the 22\arcmin~diameter focal plane. Each IFU bundle consists of 448 fibers with a diameter of 1.5\arcsec, covering a field of view of 51\arcsec$\times$51\arcsec.
The survey started in January 2017 and was completed in August 2024, with a typical image quality of $\sim$~1.8\arcsec~(full width at half maximum; FWHM). 
The survey design, data reduction procedures, and emission-line detection algorithms are described in \citet{Gebhardt2021}.

The second HETDEX Data Release (HDR2) source catalog (200,000 sources with 51,863 $z\sim2.5$ LAEs) is available in \cite{Cooper2023}. The HDR4 AGN catalog (survey 72\% completed) consists of 15,940 AGN down to $g\sim25$ \citep{Liu2022a, Liu2025}. 
In this study, we extract HETDEX spectra as one of the spectroscopic epochs used for multi-epoch comparisons, enabling a systematic search for \clagn candidates when paired with external survey spectra (SDSS, DESI, and LAMOST).

\paragraph{Flux calibration and systematic uncertainties.}
\rv{The absolute flux calibration of HETDEX spectra is non-trivial due to the fixed-altitude design of the Hobby--Eberly Telescope and the fiber-fed nature of the VIRUS spectrographs \citep{Gebhardt2021}.
HETDEX therefore adopts a field-based calibration strategy that derives a global system response from ensembles of stars distributed across the full field of view (typically $\sim$30 stars per exposure), rather than relying on a single spectrophotometric standard star.
}

\rv{This ensemble-based approach significantly reduces sensitivity to any individual star and averages over local extraction errors, astrometric uncertainties, and seeing variations.
Internal cross-checks between HETDEX and SDSS stellar spectra demonstrate that the relative flux calibration is stable at the $\sim$5--10\% level across the VIRUS wavelength range, with the dominant uncertainties arising from extraction systematics rather than temporal calibration drift.}

\rv{In this work, potential calibration-related artifacts are further mitigated by conservative statistical thresholds on emission-line variability and by visual inspections that examine contemporaneous photometric light curves.
Candidates whose apparent variability is inconsistent with broadband behavior are excluded from the final sample.}
% -----------------------------------------------------------------------------------
\subsection{Parent Sample Construction} \label{sec_parent}

\rv{
The parent sample is constructed through a two-step procedure.
We first define the search channels used to identify candidate multi-epoch spectroscopic pairs, followed by a uniform extraction of HETDEX spectra at fixed sky positions.
This design ensures that sample construction and spectral extraction are clearly separated from the subsequent \clagn selection criteria.
}

\subsubsection{Forward and Reverse Searches} \label{sec_searches}
\rv{To construct a comprehensive parent sample for \clagn identification, we adopt two complementary and symmetric search channels, referred to as the \emph{forward} and \emph{reverse} searches.}

\paragraph{\emph{forward}}
\rv{In the forward search, we start from AGN catalogs compiled by external spectroscopic surveys, including SDSS DR16Q \citep{Lyke2020}, DESI DR1 \citep{desi2025}, and LAMOST-based AGN catalogs \citep{Jin2023}, and extract the corresponding HETDEX spectra at the recorded source positions.
This approach enables a uniform spectral extraction for all forward-search targets, independent of whether the source is classified as an AGN in the HETDEX epoch. In particular, no requirement is imposed that the source be classified as an AGN in both epochs, allowing transitions between AGN-like and galaxy-like spectral states to be identified.
}

\paragraph{\emph{reverse}}
\rv{Conversely, in the reverse search, we begin with AGN identified in the HETDEX HDR5 AGN catalog and retrieve available spectra from external surveys (SDSS, DESI, and LAMOST) as additional spectroscopic epochs. This channel does not require the source to be classified as an AGN in the external survey epoch, allowing spectral state changes to be identified without assuming a prior which epoch corresponds to the active or inactive phase. Together with the forward search, this ensures that cases in which only one epoch appears AGN-like are not systematically excluded.
}

\rv{Combining both channels ensures a symmetric and unbiased exploration of spectral variability across surveys with different selection functions and observational epochs.
After removing duplicates across surveys, the parent sample consists of 10,752 unique sources.
SDSS provides the primary contribution with 7,865 sources, while DESI and LAMOST add additional unique sources not covered by SDSS, contributing 2,869 and 18 sources, respectively.
Each survey includes both \emph{forward} and \emph{reverse} search components.
}

% .................................................................................
\subsubsection{HETDEX Spectral Extraction} \label{sec_extract}
%+++++++++++++++++++++++++++++++++++++++++++++++++++++++++++++++++++++
\begin{figure*}[htbp]
\centering
\includegraphics[width=\textwidth]{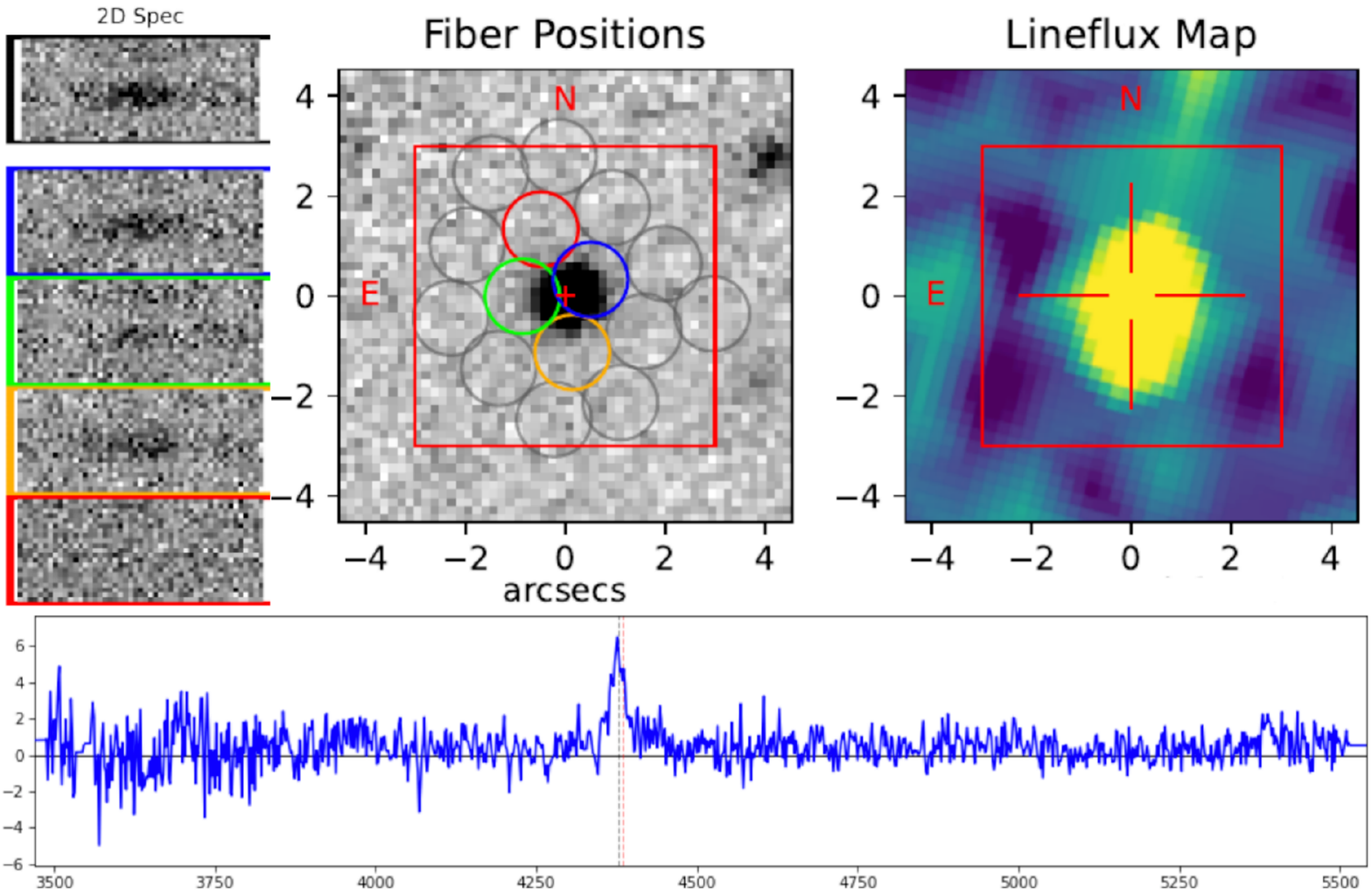}
\caption{An overview of representative HETDEX IFU data used in this work.
The top panels show, from left to right, the two-dimensional spectrum, the fiber positions overlaid on a broadband optical image (HSC $r$-band), and the corresponding line flux map for an example source. Colored circles mark representative fibers contributing to the PSF-weighted spectral extraction. The bottom panel shows the resulting one-dimensional spectrum constructed from a PSF-weighted sum of fibers. This figure illustrates the basic data structure of HETDEX observations and provides context for the extraction strategy adopted in this study.}
\label{f_fiber}
\end{figure*}
%+++++++++++++++++++++++++++++++++++++++++++++++++++++++++++++++++++++

\rv{HETDEX enables forced spectral extraction at pre-defined sky coordinates, allowing spectra to be extracted at the exact positions of known sources rather than relying on catalog-level source detections. As a result, the HETDEX epoch is examined irrespective of whether the source would be classified as an AGN, a normal galaxy, or a non-detection in that epoch.
This extraction is applied to all HETDEX spectra used in this work and is particularly critical for the \emph{forward} search, where targets are defined by external spectroscopic AGN catalogs. This approach is essential for robust multi-epoch spectral comparisons, as even sub-arcsecond positional offsets in fiber-based observations can lead to large and non-linear changes in measured emission-line fluxes. Figure~\ref{f_fiber} illustrates the point-spread-function (PSF)--weighted spectral extraction procedure and its sensitivity to fiber positioning.}

HETDEX spectra are extracted only when the source position is covered by at least one HETDEX fiber within 1~arcsec. We further require the aperture correction factor to be greater than 0.6 to exclude edge fibers with large flux uncertainties. The aperture correction is defined as the fractional fiber coverage of a $r=3\farcs5$ aperture centered on the target position, with smaller values corresponding to poorer coverage and larger flux corrections. In addition, observations with image quality worse than $\mathrm{FWHM} > 2\farcs5$ are excluded.

% -----------------------------------------------------------------------------------
\subsection{Candidate Identification} \label{sec_selection}

We identify \clagn candidates using a multi-channel selection strategy based on spectral line variability between the SDSS epoch and the HETDEX epoch(s). The goal is to construct an inclusive but physically motivated parent sample that captures both high-significance line-flux transitions and visually compelling cases that may be affected by measurement uncertainties.

\paragraph{Channel A: Flux-ratio–based selection.}
Our primary and most conservative selection follows exactly the criterion adopted by \citet{Zeltyn2024}. For a given source, we require that at least one emission line satisfies
\begin{equation}
    \frac{F_{\mathrm{bright}}}{F_{\mathrm{dim}}} - \sigma\!\left(\frac{F_{\mathrm{bright}}}{F_{\mathrm{dim}}}\right) \geq 2,
\end{equation}
where $F_{\mathrm{bright}}$ and $F_{\mathrm{dim}}$ denote the line fluxes measured in the brighter and fainter epochs, respectively. This channel yields 1,419 \clagn candidates. 

\paragraph{Channel B: S/N-based complementary selection.}
The flux-ratio criterion in Channel~A implicitly requires that the relevant emission line be meaningfully detected and modeled in both spectral epochs. 
However, in our fitting pipeline, we intentionally treat severely noise-dominated cases as non-detections. 
Specifically, when the pixel-level signal-to-noise ratio within the fitting window satisfies ${\rm SNR}_{\rm pix} < 1$, the multi-Gaussian line model is forced to degenerate into a flat (constant) model, indicating the absence of statistically significant line constraints. In such cases, all derived line parameters, including flux and ${\rm SNR}_{\rm line} \equiv F/\sigma_F$, are set to zero by construction, as illustrated by the \sio~and \ovi~fits in Figure~\ref{f_specfit}.

As a result, sources exhibiting a genuine appearance or disappearance of an emission line between epochs may fail the strict flux-ratio requirement in Channel~A, not because the transition is unphysical, but because one epoch is formally classified as a non-detection. 
To recover such cases, we introduce a complementary S/N-based selection channel that requires ${\rm SNR}_{\rm line} > 10$ in one epoch and ${\rm SNR}_{\rm pix} < 1$ in the other. 
This channel is designed to explicitly capture sources in which an emission line is robustly detected in one epoch but effectively absent in the other due to noise-dominated measurements. 
This complementary channel contributes 91 additional candidates.

\paragraph{Channel C: Literature-confirmed sources.}
Finally, we include a small number of \clagn candidates previously reported in the literature that are recovered in our data but do not satisfy the above automated criteria. This channel adds 9 candidates and serves as a completeness check rather than a statistically independent selection.

\paragraph{Final candidate pool.}
The union of the three channels results in a total of 1,519 \clagn candidates, which defines the parent sample used in the subsequent analysis. Where relevant, we explicitly distinguish between candidates selected by the conservative flux-ratio criterion (Channel A) and those identified through the auxiliary channels.

\subsection{Visual Inspections (VIs)} \label{sec_vi}

Visual inspections (VIs) are performed to identify candidates whose apparent line-flux variations are driven by flux calibration issues or spectral modeling limitations rather than genuine emission-line transitions. Such cases are not always efficiently captured by automated selection alone and are therefore examined through complementary checks on photometric consistency and spectral line profiles.

% --------------------------------------------------
\subsubsection{VI: Photometric Consistency} \label{sec_vi_phot}

In some instances, the reliability of spectral flux measurements cannot be independently validated due to insufficient or ambiguous photometric coverage. We therefore visually inspect the concurrent light curves of each candidate to assess whether the spectral flux levels inferred from HETDEX and external surveys are consistent with the temporal photometric behavior.

Pseudo-$g$ magnitudes are derived by convolving both HETDEX and external spectra with the SDSS $g$-band transmission curve. Candidates are rejected if the derived pseudo-$g$ magnitudes exhibit significant discrepancies with contemporaneous photometric measurements, or if the available photometric data are too sparse to meaningfully validate the spectral epochs.
An example of a candidate excluded due to insufficient photometric coverage is shown in the top panel of Figure~\ref{f_vi_rej}, where the lack of overlapping light-curve data prevents a reliable assessment of the spectral flux calibration.

We collect archival light curves from the $V$-band observations of the Catalina Real-Time Transient Survey (CRTS; \citealt{Drake2009}), the $g_{\mathrm{PS1}}$-band photometry from Pan-STARRS Data Release 2 \citep[\rv{PanSTARRS};][]{Chambers2016,Flewelling2020}, and the Zwicky Transient Facility (ZTF; \citealt{Bellm2019}). PS1 and ZTF are particularly useful for validating the SDSS and HETDEX spectral epochs, respectively, owing to their temporal overlap. Although the $g$-band filters differ slightly among surveys, the expected offsets are negligible ($<0.01$~mag) for AGNs at the typical redshift of this work, and no filter conversions are applied during VI.

% --------------------------------------------------
\subsubsection{VI:  Spectral Line Profiles and Absorption} \label{sec_vi_spec}

Visual inspection of the two spectral epochs is also used to identify cases where apparent variability is driven by spectral modeling limitations rather than genuine line transitions. In low signal-to-noise regimes, automated fitting can produce unstable or misleading line-flux measurements, particularly when one epoch is strongly noise-dominated. In addition, complex line morphologies affected by broad absorption features may give rise to large apparent flux changes that do not reflect intrinsic emission-line variability.

Representative examples are shown in Figure~\ref{f_vi_rej}. The middle panel illustrates a candidate for which an additional DESI spectrum is excluded due to inconsistencies with both the SDSS and HETDEX spectral epochs, indicating that the apparent variability is not robust across independent spectroscopic datasets. The bottom panel shows a source in which changes in absorption features dominate the apparent line variability, leading to significant differences in the fitted emission-line fluxes despite broadly consistent photometric behavior. Such cases are rejected, as the observed variability cannot be confidently attributed to changes in the ionizing continuum.

% --------------------------------------------------
After applying the visual inspection procedures described above, candidates whose variability signatures are inconsistent with either photometric behavior or genuine emission-line transitions are removed, yielding a final sample of \nclagn \clagn.

% -----------------------------------------------------------------------------------
\subsection{The final \clagn sample} \label{sec_clagn_sample}

The final \clagn sample consists of \nclagn\ sources, listed in Table~\ref{t_clagn_thiswork}. Of these, \nclagnnew\ are newly identified in this work, supplementing our compilation of \nlit \clagn from 39 references (Appendix~\ref{sec_clagn_lit}). Although nine previously reported \clagn from the literature fall within our parent sample, only two are recovered in the final emission-line variability (EVA) sample defined in this work.
The remaining sources do not show clear or reproducible emission-line transitions between the HETDEX and external spectral epochs available to us. This outcome does not imply a contradiction with previous classifications, but reflects differences in data coverage, temporal baselines, and selection methodologies. In particular, some literature \clagn are identified based on continuum variability or spectral changes occurring outside the epochs probed in our analysis, and therefore cannot be robustly recovered within our EVA-based framework.

%+++++++++++++++++++++++++++++++++++++++++++++++++++++++++++++++++++++
\begin{deluxetable*}{ccccccccccc}
\tablecaption{The final \clagn sample identified in this work (ordered by right ascension).\label{t_clagn_thiswork}}
\tablehead{
\colhead{(1)} &
\colhead{(2)} &
\colhead{(3)} &
\colhead{(4)} &
\colhead{(5)} &
\colhead{(6)} &
\colhead{(7)} &
\colhead{(8)} &
\colhead{(9)} &
\colhead{(10)} &
\colhead{(11)}\\
\colhead{Name} &
\colhead{Redshift} &
\colhead{Transition} &
\colhead{Line} &
\colhead{$F_{\rm bright}/F_{\rm dim}$} &
\colhead{$\sigma_{\rm ratio}$} &
\colhead{Method} &
\colhead{$g$} &
\colhead{$g_{\rm er}$} &
\colhead{MJD$_{\rm H}$} &
\colhead{MJD$_{\rm E}$}
}
\startdata
J004920.93+020224.5 & 0.966 & dimming     & CIII & 2.334 & 0.143 & lineratio & 20.746 & 0.009 & 59139 & 57278 \\
J005118.45-013929.7 & 0.987 & brightening & CIII & 3.567 & 0.620 & lineratio & 20.524 & 0.007 & 60311 & 56902 \\
J005336.53-013329.6 & 0.411 & dimming     & MgII & 2.080 & 0.056 & lineratio & 18.640 & 0.002 & 59464 & 58074 \\
J005613.47-001111.2 & 0.425 & dimming     & MgII & 2.375 & 0.298 & lineratio & 20.831 & 0.010 & 58452 & 55455 \\
J005717.64+005158.1 & 1.032 & brightening & CIII & 4.930 & 0.466 & lineratio & 19.860 & 0.004 & 58788 & 58097 \\
\ldots & \ldots & \ldots & \ldots & \ldots & \ldots & \ldots & \ldots & \ldots & \ldots & \ldots \\
\enddata
\tablecomments{
(1) IAU name in JHHMMSS.ss$\pm$DDMMSS.s format (J2000 equinox). 
(2) Spectroscopic redshift. 
(3) Transition type (``brightening'' or ``dimming'').
(4) Dominant emission line associated with the transition.
(5) Line-flux ratio between the bright and dim epochs, $F_{\rm bright}/F_{\rm dim}$.
(6) Uncertainty of Column~(5).
(7) Selection channel: \texttt{lineratio} (flux-ratio criterion), \texttt{snr} (S/N complementary criterion), or \texttt{lit} (previously reported in the literature).
(8) Pseudo-$g$ AB magnitude computed by integrating the representative HETDEX spectrum through the SDSS $g$-band transmission curve.
(9) Uncertainty of Column~(8); values below 0.01 mag are reported to three decimal places to avoid rounding to zero.
(10) Modified Julian Date (MJD) of the representative HETDEX spectral epoch used for the reported measurements.
(11) MJD of the external spectral epoch paired with the HETDEX epoch in computing the line-flux ratio.
Only a portion of the table is shown here; the full table is available in machine-readable form.
}
\end{deluxetable*}
%+++++++++++++++++++++++++++++++++++++++++++++++++++++++++++++++++++++

% ====================================================================================
\section{Results and Discussion} \label{sec_result}

%+++++++++++++++++++++++++++++++++++++++++++++++++++++++++++++++++++++
\begin{figure}[htbp]
\centering
\includegraphics[width=0.49\textwidth]{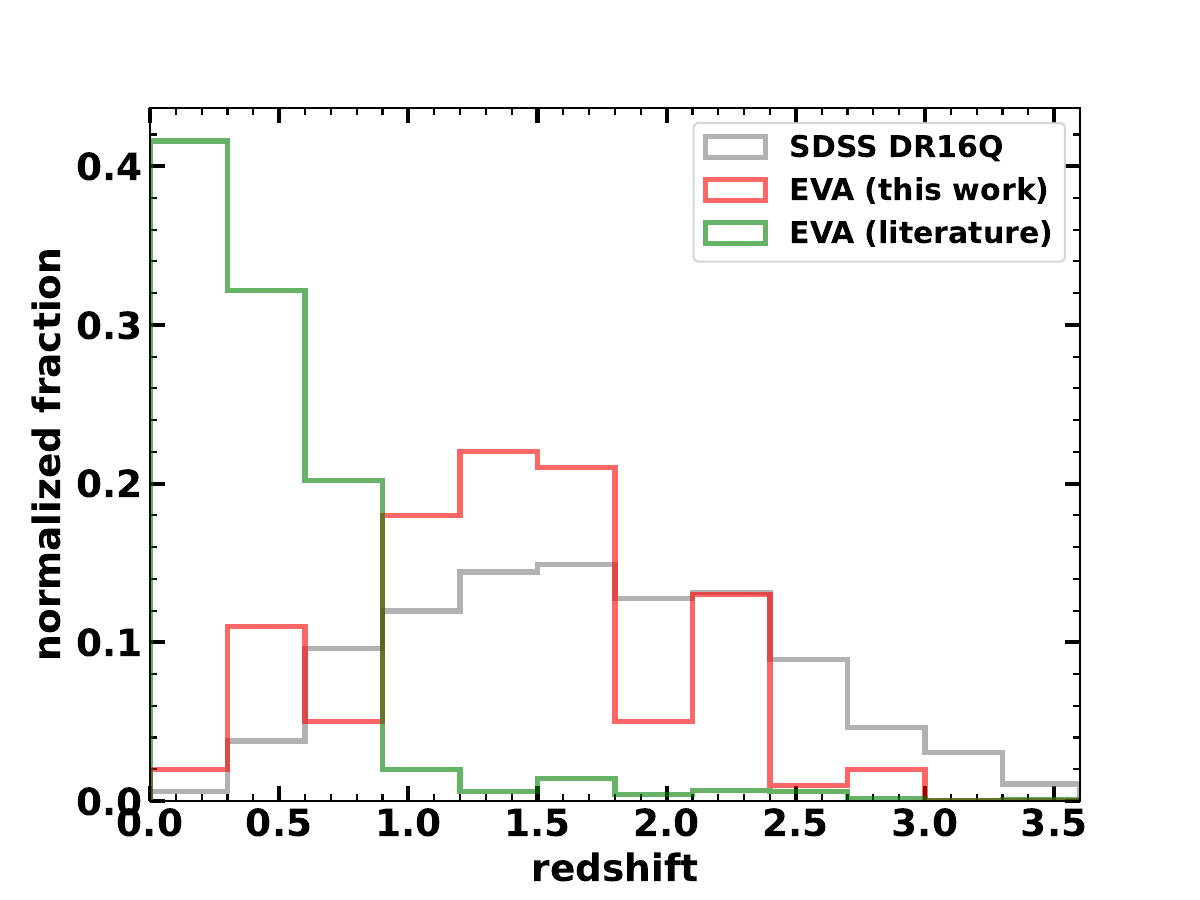}
\caption{Redshift distributions of the \clagn of this work (red) and in the literature (green). The redshift distribution of the SDSS DR16Q sample is shown in gray.}
\label{f_redshift}
\end{figure}
%+++++++++++++++++++++++++++++++++++++++++++++++++++++++++++++++++++++

Using the final \clagn sample defined in Section~\ref{sec_clagn_sample},
we now present the statistical properties of the emission-line variable AGN
identified in this work.

Figure~\ref{f_redshift} compares the redshift distributions of the \clagn sample
identified in this work with those compiled from the literature, together with
the SDSS DR16Q population.
Owing to the untargeted nature of HETDEX, our sample is not biased toward
favoring brighter or lower-redshift sources at the second epoch, and therefore
does not introduce additional redshift-dependent selection effects relative to
the parent AGN sample.
As a consequence, the \clagn identified in this work extends to systematically
higher redshifts than most previously reported emission-line variable AGN.

% -----------------------------------------------------------------------------------
\subsection{Duty Cycle} \label{sec_dutycycle}

If one interprets the fraction of \clagn identified under our operational
definition as an estimate of the duty cycle of significant spectroscopic
variability, then for our sample this fraction is approximately
$f_{\rm \clagn} = \nclagno/\ntot \sim 0.9\%$ of the AGN net lifetime $t_{\rm net}$.
This value is comparable to, though somewhat higher than, incidence fractions
reported in lower-redshift studies (e.g., the $\sim0.4\%$ fraction reported by
\citealt{Zeltyn2024}), which may reflect differences in survey strategy,
temporal baseline between spectroscopic epochs, redshift coverage, and
selection sensitivity.

It is important to emphasize that this fraction should be regarded as a
lower limit to the true incidence of dramatic emission-line variability.
Some genuine \clagn events may not be recovered owing to incomplete temporal sampling of contemporaneous photometric light curves used in the visual inspection, or line-dependent selection effects inherent to fixed wavelength coverage. As such, the inferred fraction is not
intended as a precise measurement of a universal duty cycle, but rather as an
order-of-magnitude estimate within the context of our operational definition.

Multiplying this \clagn fraction by the net quasar lifetime,
$t_{\rm net} \sim 10^{6}$--$10^{8}$ yr \citep{Martini2001}, yields a total
timescale of significant emission-line variability of
$t_{\rm net,\clagn} \gtrsim 10^{4}$--$10^{6}$ yr. This estimate does not constrain
the duration of individual transitions, which we investigate in the following
subsection.

% -----------------------------------------------------------------------------------
\subsection{Time-scale} \label{sec_timescale}

%+++++++++++++++++++++++++++++++++++++++++++++++++++++++++++++++++++++
\begin{figure}[htbp]
\centering
\includegraphics[width=\linewidth]{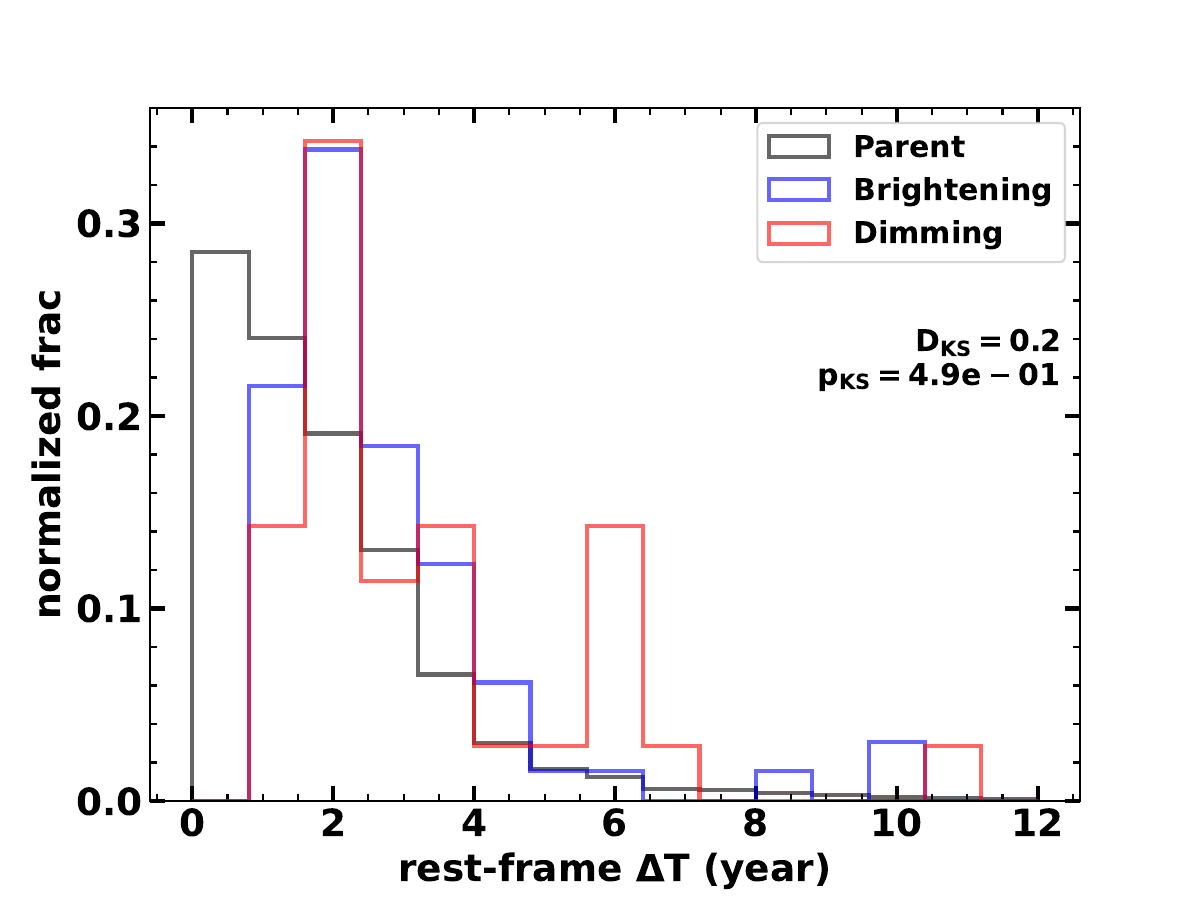}
\caption{Distribution of the rest-frame time interval between the \ntot parent AGN sample constructed from SDSS, DESI, and LAMOST spectroscopic catalogs with HETDEX spectral coverage (black). The Kolmogorov–Smirnov (KS) test parameters between the brightening and the dimming sample is $D_{\rm KS}=0.2$ and $p_{\rm KS}=0.49$.}
\label{f_timescale}
\end{figure}
%+++++++++++++++++++++++++++++++++++++++++++++++++++++++++++++++++++++

The rest-frame time interval between the external spectroscopic epoch and the
HETDEX epoch provides an observational upper limit on the episodic transition
timescale of our \clagn sample. Figure~\ref{f_timescale} presents the
distributions of these upper limits for the brightening and dimming \clagn
subsamples. The brightening subset exhibits a median rest-frame timescale of
$\Delta T \sim 2.2$ yr, while dimming transitions are marginally longer
($\Delta T \sim 2.6$ yr). This difference is statistically insignificant
($D_{\rm KS}=0.2$, $p_{\rm KS}=0.49$), indicating that the transition timescales
of brightening and dimming events are broadly consistent within the limits of
our temporal sampling.

The absence of a statistically significant asymmetry between brightening and
dimming timescales suggests that the observed transitions are unlikely to be
dominated by strongly time-asymmetric processes, such as tidal disruption
events or nuclear supernovae, which typically exhibit pronounced differences
between rise and decay timescales. Instead, the observed symmetry is consistent
with scenarios involving recurrent or stochastic changes in the accretion
state, in agreement with the findings of \citet{WangS2025}, who reported
comparable brightening and dimming timescales for recurrent \clagno.

Based on the distribution of rest-frame time intervals, we infer a
characteristic episodic transition timescale of
$t_{\rm episodic,\clagn} \sim 1$--$10$ yr, consistent with previous studies
\citep[e.g.,][]{Lu2025}. Combining this estimate with the duty-cycle constraint
derived in Section~\ref{sec_dutycycle}, a typical quasar may experience up to
\[
N \lesssim \frac{t_{\rm net,\clagn}}{t_{\rm episodic,\clagn}} \sim 10^{3-5}
\]
episodes of pronounced spectroscopic variability over its lifetime. This value
should be regarded as an order-of-magnitude upper limit, as the observed
\clagn fraction is subject to selection effects, and not every identified
transition necessarily corresponds to the onset or termination of a sustained
active phase.

% -----------------------------------------------------------------------------------
\subsection{Line Variation} \label{sec_lines}

%+++++++++++++++++++++++++++++++++++++++++++++++++++++++++++++++++++++
\begin{figure*}[htbp]
\centering
\includegraphics[width=0.495\linewidth]{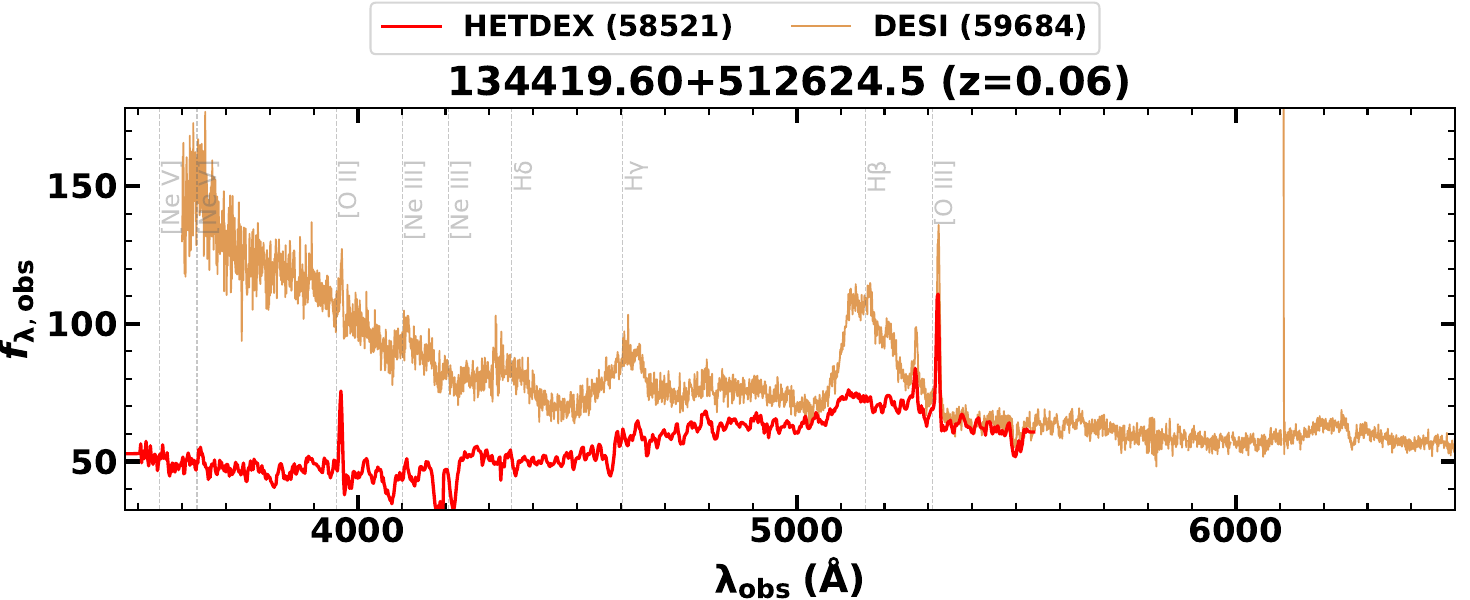}
\includegraphics[width=0.495\linewidth]{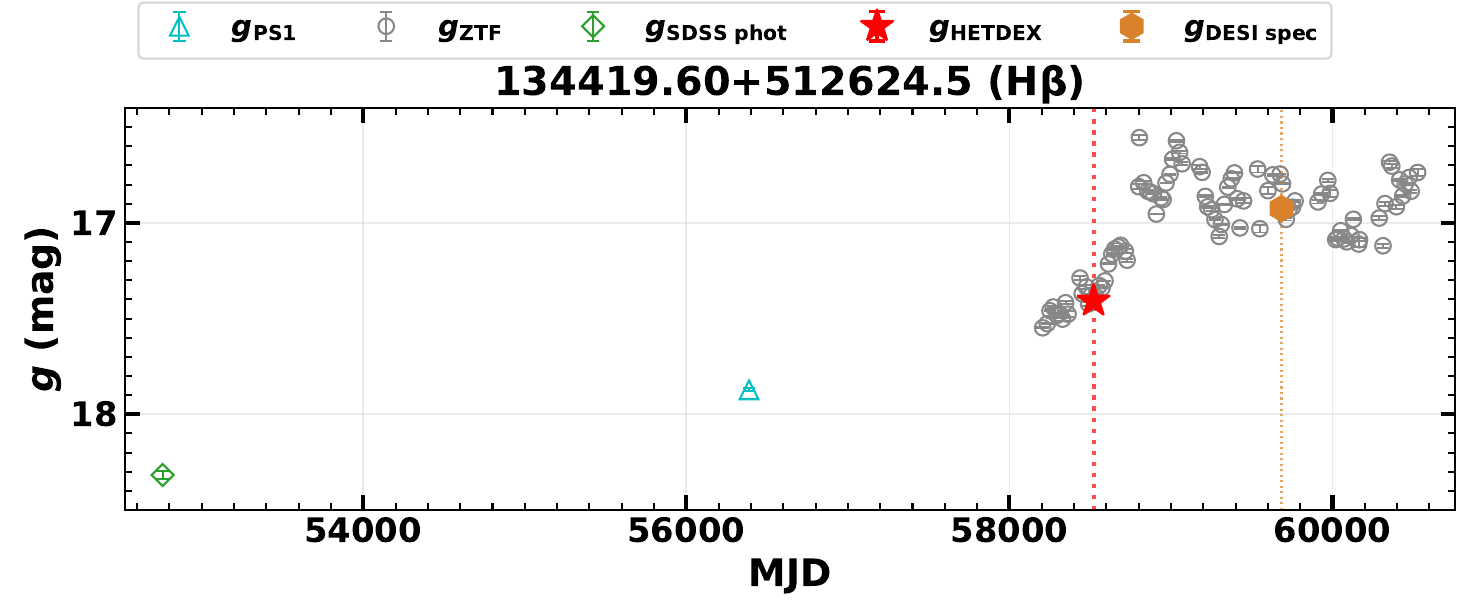}\\
\includegraphics[width=0.495\linewidth]{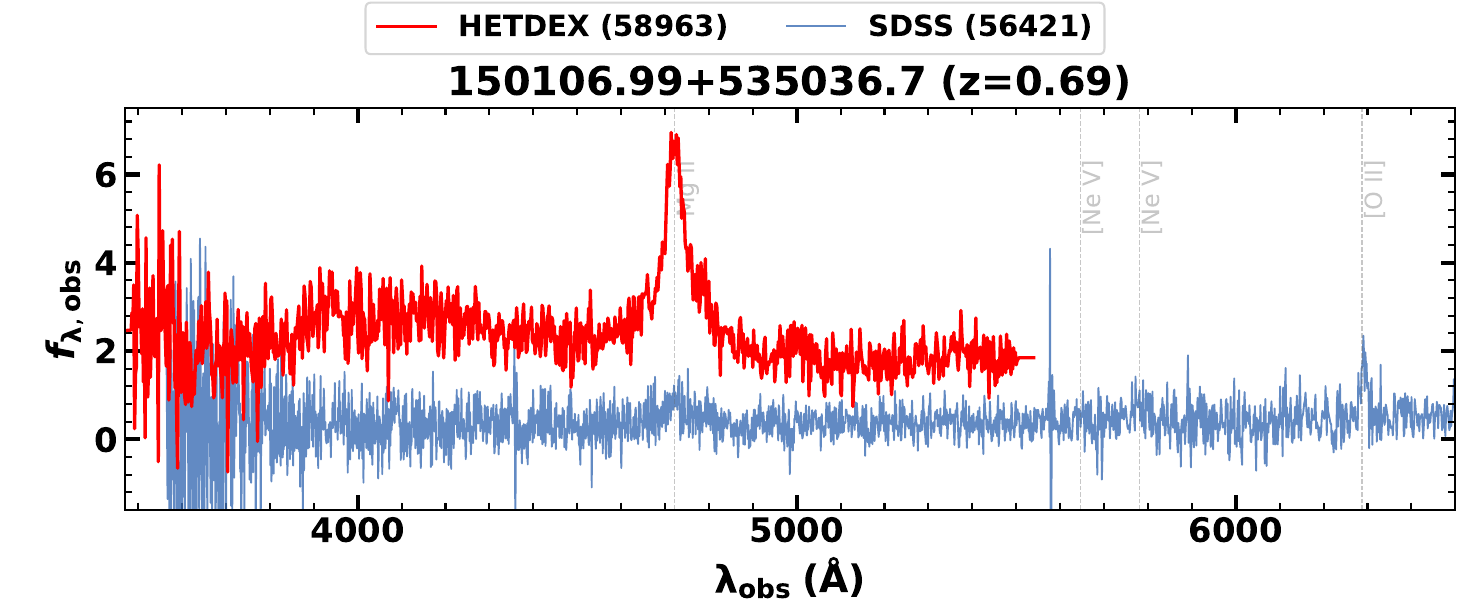}
\includegraphics[width=0.495\linewidth]{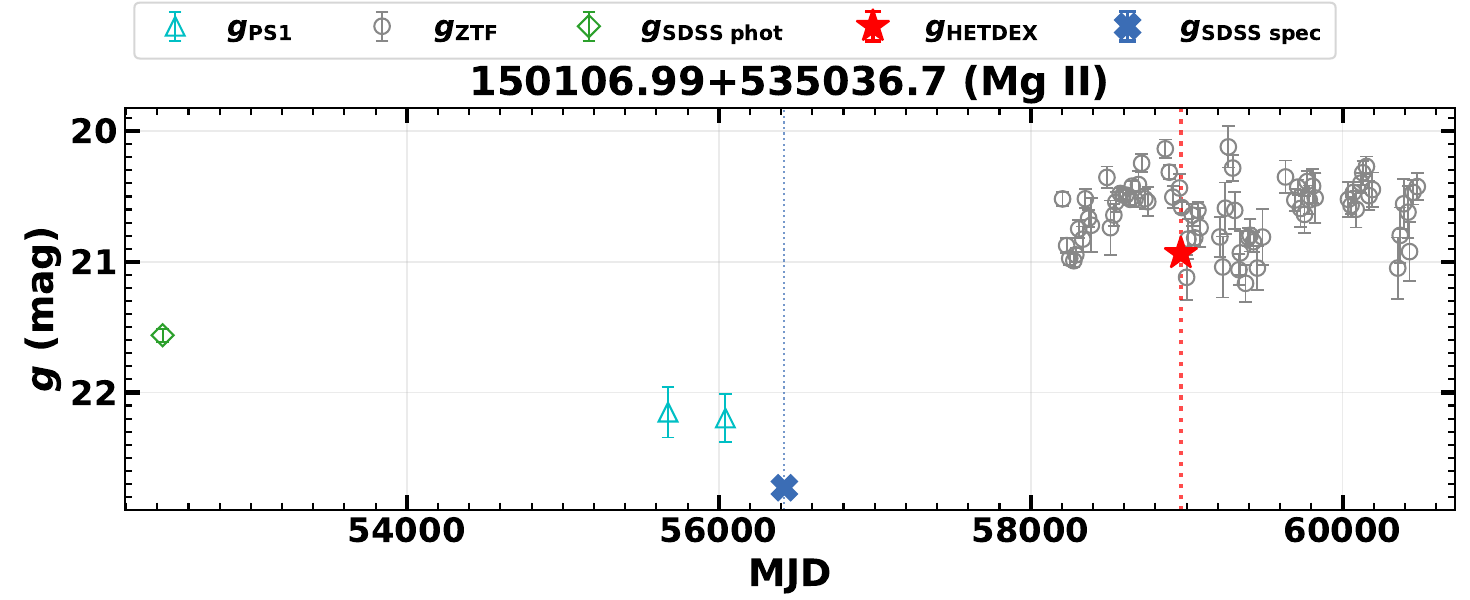}\\
\includegraphics[width=0.495\linewidth]{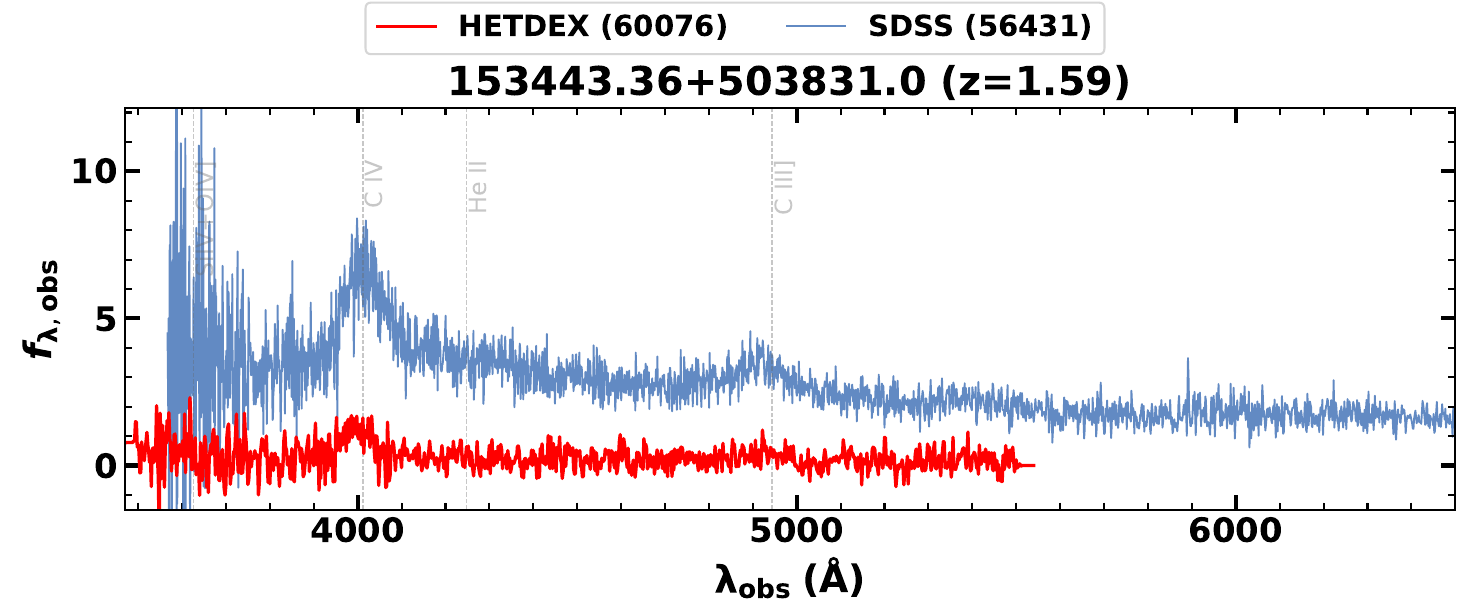}
\includegraphics[width=0.495\linewidth]{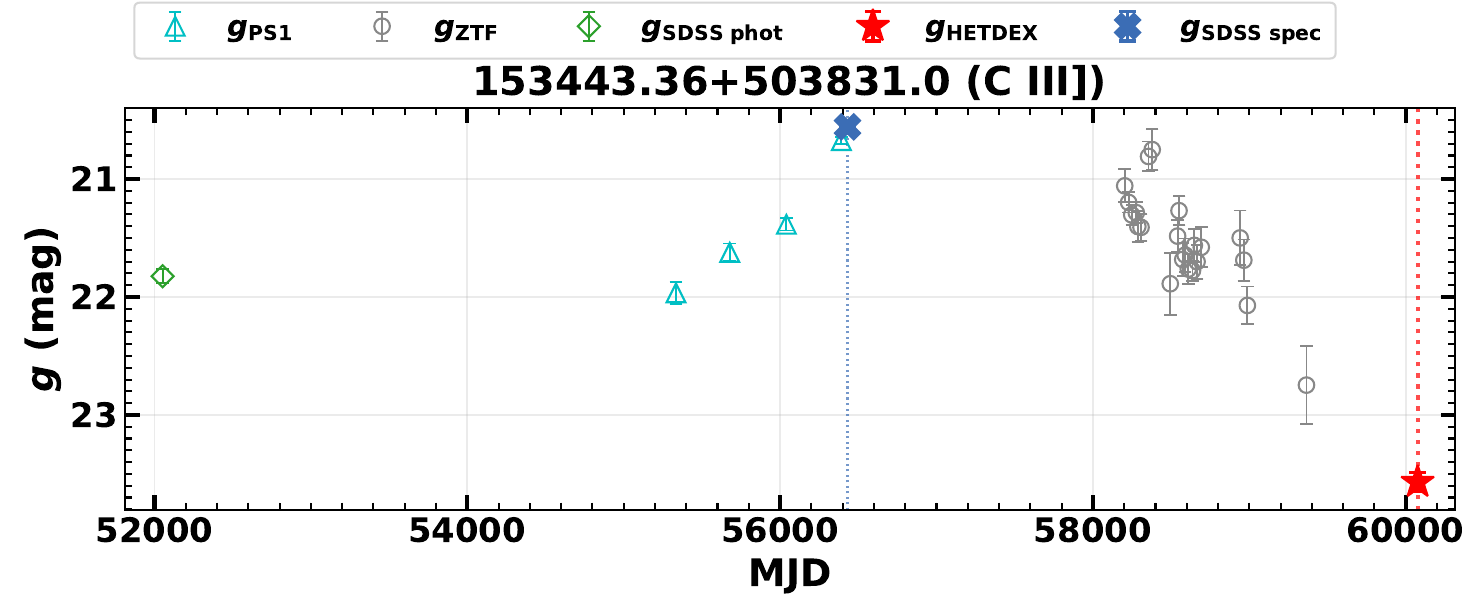}\\
\includegraphics[width=0.495\linewidth]{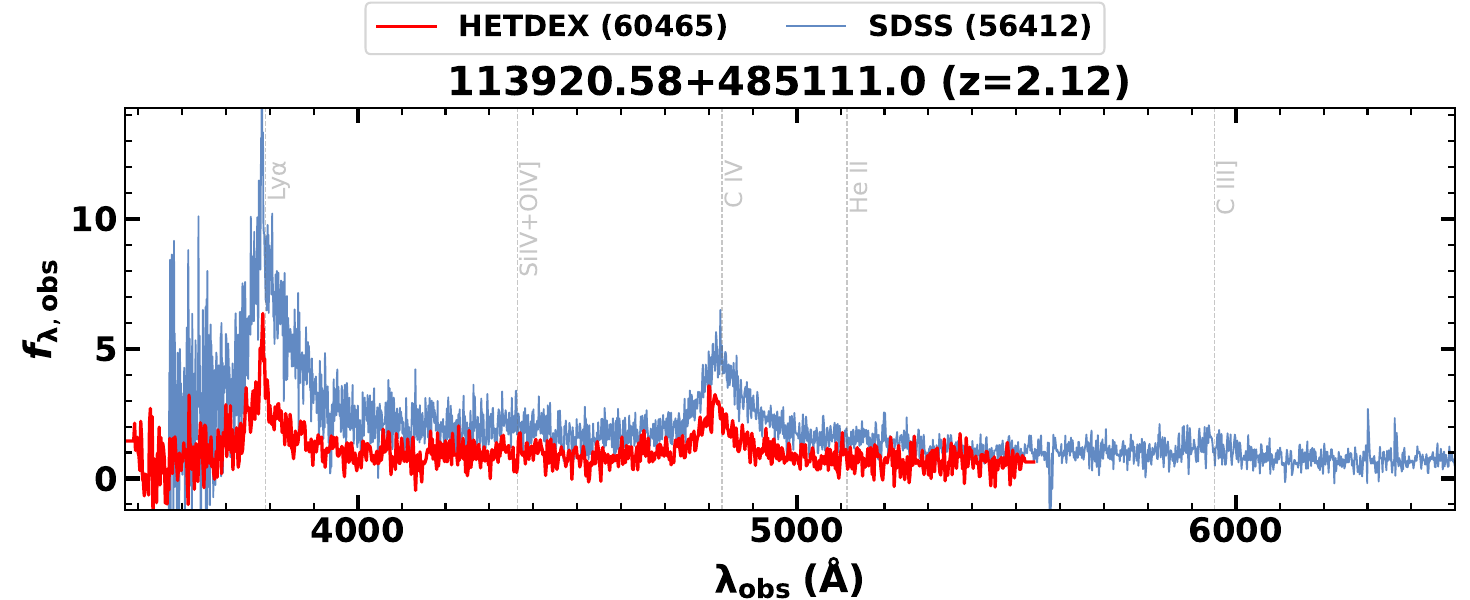}
\includegraphics[width=0.495\linewidth]{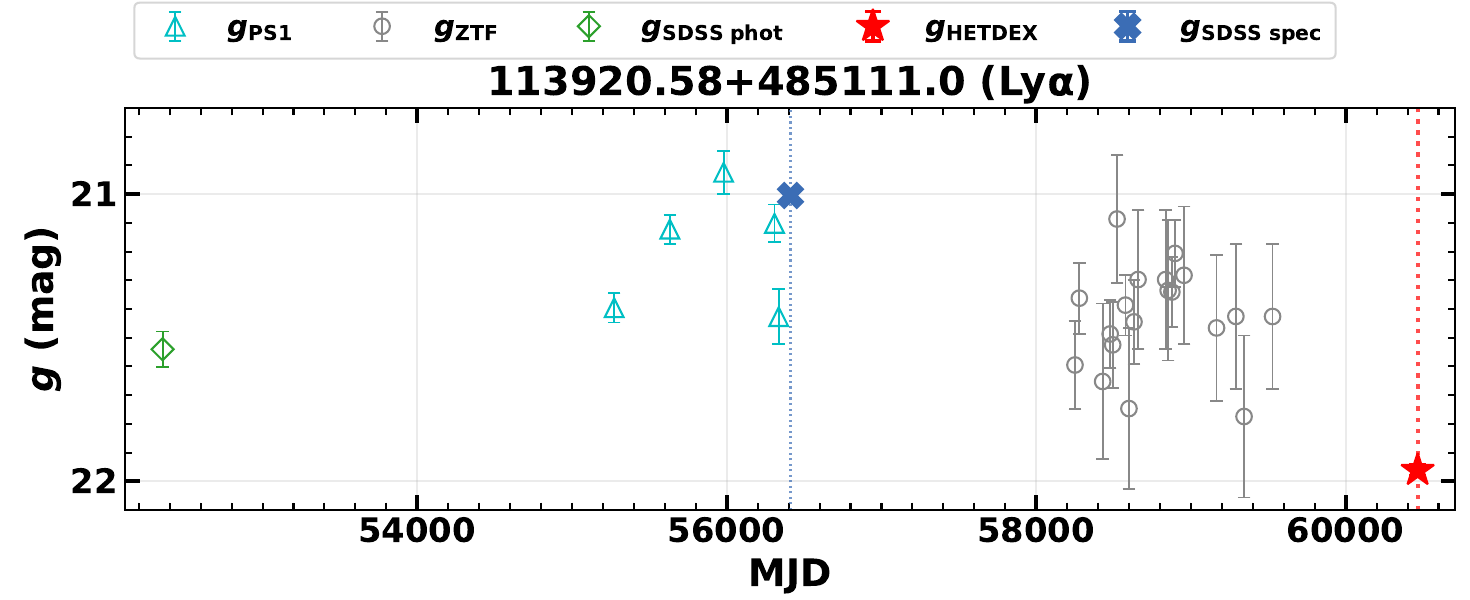}\\
\caption{Representative examples of emission-line variable \clagn identified in this work.
From top to bottom, the \clagn cases are associated with variability in \hb, \mgii, \ciii, and \lya, respectively.
Left panels: Comparison between the HETDEX spectra (red) and the external spectra (SDSS or DESI; blue/orange), shown in the observed-frame wavelength. Prominent emission lines are marked for reference.
Right panels: Multi-epoch $g$-band light curves compiled from SDSS imaging, PS1, and ZTF, together with pseudo-$g$ magnitudes derived from the spectra. The pseudo-$g$ magnitudes are computed by integrating the SDSS and HETDEX spectra through the SDSS $g$-band transmission curve, and are shown as solid blue squares (SDSS spectra) and solid red stars (HETDEX spectra), respectively. Vertical dotted lines indicate the epochs of the corresponding spectroscopic observations.
These examples illustrate that the identified \clagn is consistently reflected in both spectroscopic emission-line changes and contemporaneous photometric behavior.
}
\label{f_examples}
\end{figure*}
%+++++++++++++++++++++++++++++++++++++++++++++++++++++++++++++++++++++

Figure~\ref{f_examples} presents representative examples of emission-line variable \clagn identified in this work, selected to illustrate variability detected through different broad emission lines.
From top to bottom, the examples correspond to \clagn identified via variability in \hb, \mgii, \ciii, and \lya, respectively.
In each case, the left panel compares the HETDEX spectrum (red) with an external spectrum from SDSS or DESI (blue or orange), shown in the observed-frame wavelength, with prominent emission lines marked for reference.
The right panel shows the multi-epoch $g$-band light curves compiled from SDSS imaging, PS1, and ZTF, together with pseudo-$g$ magnitudes derived from the spectra by integrating the spectra through the SDSS $g$-band transmission curve. Vertical dotted lines indicate the epochs of the corresponding spectroscopic observations.
Except for the representative Ly$\alpha$ example shown in Figure~\ref{f_examples}, we do not observe clear cases of broad-to-narrow or narrow-to-broad line transitions in our sample.
Instead, the majority of \clagn identified in this work exhibit coherent changes in the overall flux or profile shape of the broad emission lines, rather than complete appearance or disappearance.

Among the \nclagn \clagn sample, emission-line variability is identified in 43, 28, 15, 2, and 12 sources based on \civ, \ciii, \mgii, \hb, and \lya, respectively.
These numbers correspond to sources in which the corresponding emission lines fall within the HETDEX spectral coverage (3500--5500\,\AA).
We do not include \ovi~in the following analysis, as the \ovi~line lies in the Ly$\alpha$ forest region, where strong absorption features and uncertain continuum placement frequently lead to unstable line-profile fitting with large $\chi^2$, preventing robust line variability identification.
 
Overall, emission-line variability is detected across \ciii, \civ, \lya, and \mgii\ when moderate but statistically significant line variations are considered.
We note that Ly$\alpha$-selected \clagn are relatively less frequent compared to \civ\ and \ciii, which may be partly related to the resonant nature of the Ly$\alpha$ line and its susceptibility to radiative transfer effects and absorption.
We note that previous studies focusing on extreme changing-look events defined by the complete appearance or disappearance of broad lines have found \lya~and \civ~CL-AGN to be rare (e.g., \GuoHX; \GuoWJ), highlighting the sensitivity of \lya~and \civ~CL-AGN incidence to the adopted definition.
Variability identified via \hb~is rare, primarily owing to the limited redshift range over which \hb~is accessible within the HETDEX wavelength coverage.
As illustrated in Figure~\ref{f_examples}, in all cases the spectroscopically identified emission-line variability is accompanied by contemporaneous photometric variability in the $g$ band, supporting the interpretation that the observed line changes are driven by intrinsic changes in the AGN continuum.

% -----------------------------------------------------------------------------------
\subsection{Brightening-to-dimming Ratio}\label{sec_btdr}

%+++++++++++++++++++++++++++++++++++++++++++++++++++++++++++++++++++++
\begin{figure}[htbp]
\centering
\includegraphics[width=\linewidth]{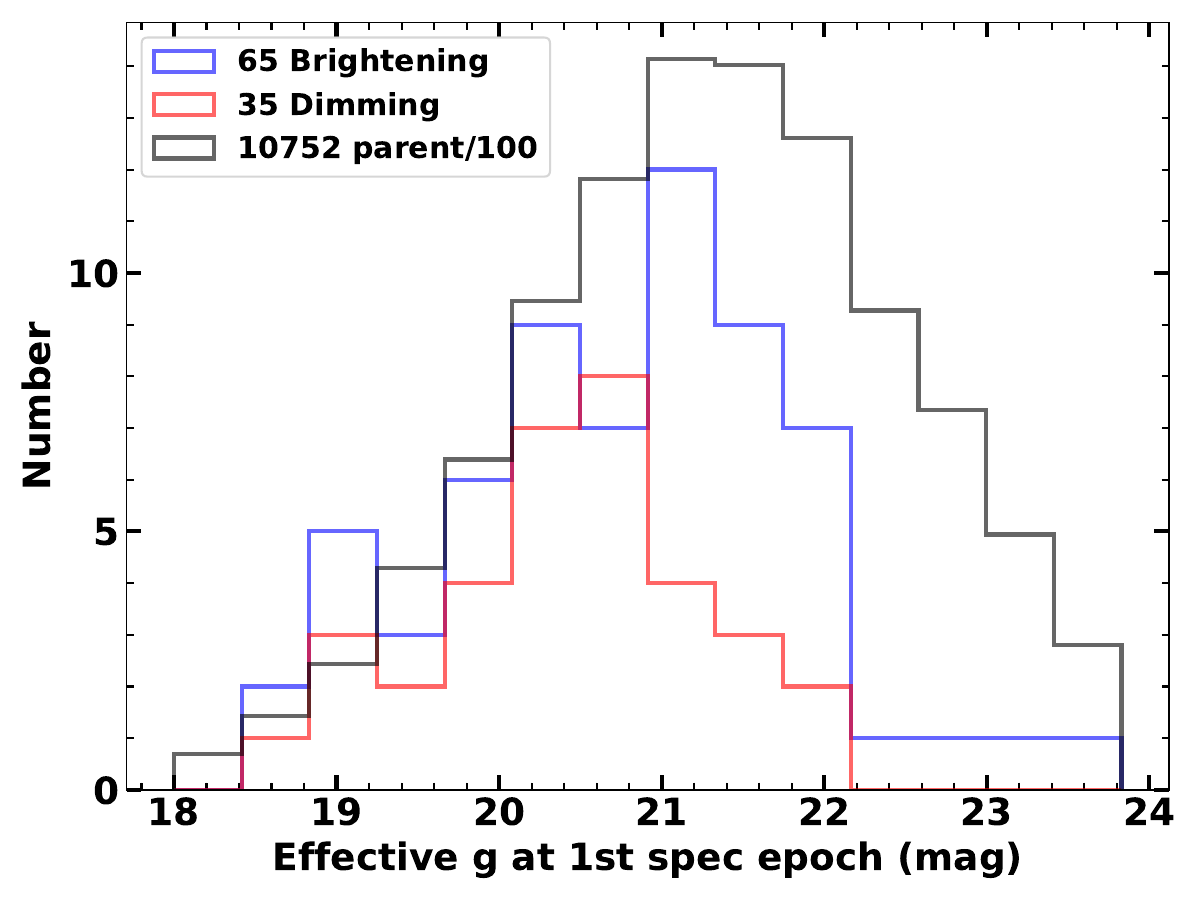}
\caption{
Distribution of the effective $g$-band magnitude at the first spectroscopic epoch for the brightening \clagn (blue), the dimming \clagn (red), and the scaled external $\times$ HETDEX parent quasar sample (black; divided by a factor of 100 for visual comparison).
}
\label{f_onoff}
\end{figure}
%+++++++++++++++++++++++++++++++++++++++++++++++++++++++++++++++++++++

The identification of \clagn is inevitably shaped by selection effects, which can introduce an asymmetry between brightening and dimming events. In many spectroscopic follow-up programs, repeat observations preferentially target quasars that are already luminous, leading to a higher detection efficiency for dimming events and a commonly reported brightening-to-dimming ratio of $\sim$ 1:2 in the literature \citep[e.g.,][]{Zeltyn2024}.

Although HETDEX itself provides spectroscopic coverage that is independent of source brightness, our \clagn sample relies on associations between HETDEX observations and external spectroscopic surveys, including both forced HETDEX spectral extraction at the positions of externally identified AGN and positional cross-matching of HETDEX AGN to available external spectra. As a consequence, the effective parent sample of \clagn is ultimately defined by the availability of external spectroscopic observations, regardless of the direction of association. Potential brightening AGN that were too faint to be included in external spectroscopic surveys at the first spectroscopic epoch are therefore excluded from our analysis, even though HETDEX itself is an untargeted survey.

An additional selection effect arises near the flux limits of the external spectroscopic surveys. Quasars with relatively faint initial magnitudes ($g \gtrsim 21$) that undergo dimming are more likely to fall below the HETDEX spectral S/N threshold and thus be lost from the sample, whereas objects that brighten remain detectable and can be identified as brightening events. This asymmetry naturally inflates the observed fraction of brightening \clagn at the faint end.

As shown in Figure~\ref{f_onoff}, this effect is clearly manifested in the distribution of effective $g$-band magnitude at the first spectroscopic epoch. Brightening events dominate at $g \gtrsim 21$, while at brighter magnitudes ($g \lesssim 21$) the numbers of brightening and dimming \clagn become comparable, with a brightening-to-dimming ratio close to unity, indicating a regime where luminosity-driven selection effects are substantially reduced. Taken together with the absence of a statistically significant asymmetry in the rest-frame timescale distributions (Figure~\ref{f_timescale}), the near-unity brightening-to-dimming ratio observed in the bright subsample is consistent with a largely symmetric physical process governing brightening and dimming events.

Considering the full sample, we obtain an overall brightening-to-dimming ratio of approximately 65:35, corresponding to $\sim$ 2:1. This moderate excess of brightening events should not be interpreted as an intrinsic transition ratio, but is largely driven by the selection effects described above. We further note that the redshift distribution of our sample peaks at $z \sim 1.5$, near cosmic noon, when AGN activity and accretion variability are expected to be more prevalent. In such a regime, the probability of observing large-amplitude spectral variability over a fixed temporal baseline may be enhanced.

Finally, we note that none of the sources in our sample exhibit a complete transition from an AGN to a non-AGN spectral state within the available spectroscopic epochs.

% -----------------------------------------------------------------------------------
\subsection{Baldwin Effect}\label{sec_baldwin}

%+++++++++++++++++++++++++++++++++++++++++++++++++++++++++++++++++++++
\begin{figure*}[htbp]
\centering
\includegraphics[width=0.49\linewidth]{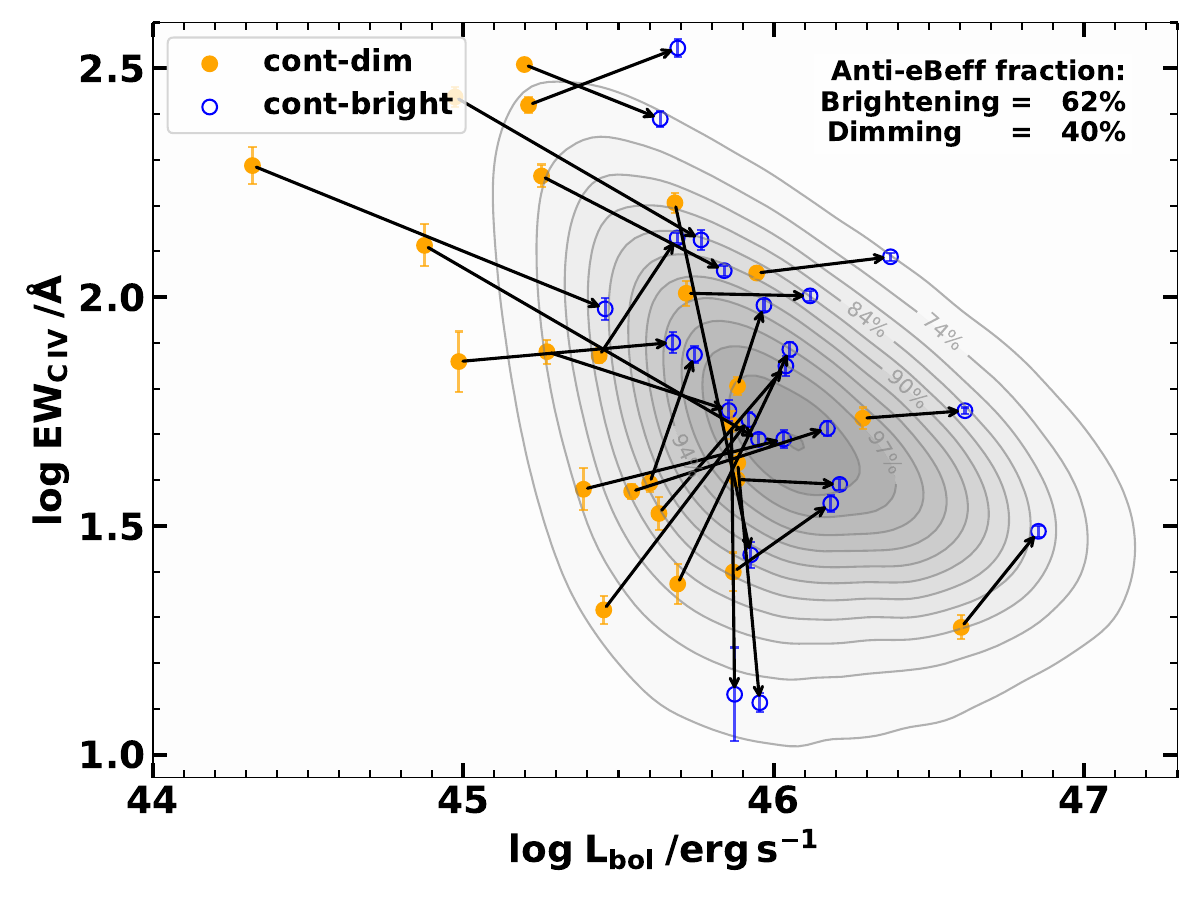}
\includegraphics[width=0.49\linewidth]{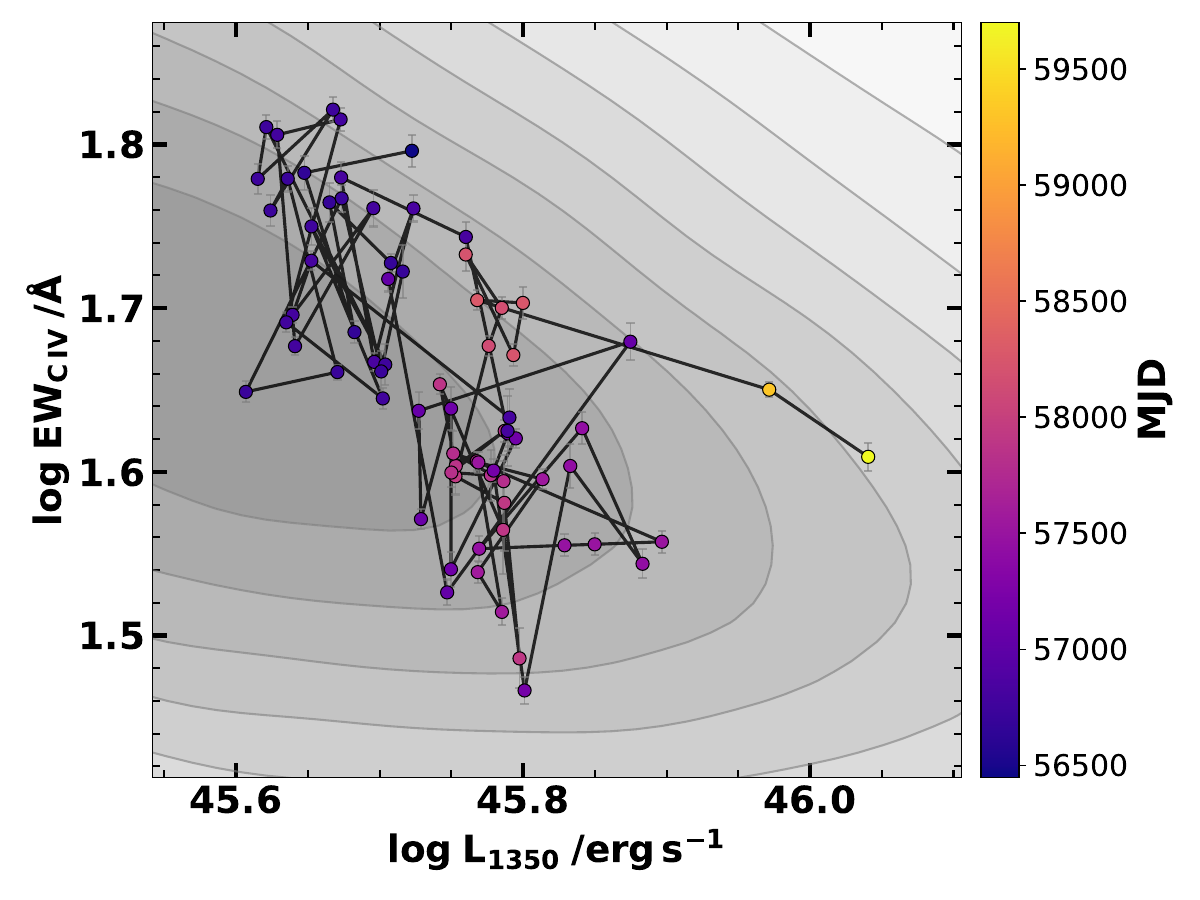}
\caption{
Baldwin effect \citep{Baldwin1977} for \civ\ in our \clagn\ sample.
\textit{Left}: Rest-frame EW of the \civ\ emission line versus bolometric luminosity ($L_{\rm bol}$) for \clagn\ sources observed at two spectral epochs (continuum-bright and continuum-dim states). Only sources with both epochs satisfying quality cuts ($\mathrm{S/N}_{\mathrm{C\,IV}} > 5$ and $\chi^2_{\mathrm{C\,IV}} < 2.5$) are shown, in order to minimize the impact of complex absorption on line fitting. Orange filled circles denote the continuum-dim state (lower $L_{\rm bol}$) and blue open circles denote the continuum-bright state (higher $L_{\rm bol}$) for each source. Grey contours represent the ensemble Baldwin effect (eBeff) measured from the SDSS DR16Q quasar sample \citep{Wu2022}. The labeled anti-eBeff fractions indicate the fraction of \clagn\ whose EW response between epochs is opposite to the expectation from the ensemble trend.
\textit{Right}: (Zoomed-in) intrinsic Baldwin evolution (iBeff) of J142308.03+522815.6, based on 73 spectroscopic epochs (71 SDSS, 1 DESI, and 1 HETDEX). Points are color-coded by observation time (MJD), illustrating a coherent, time-resolved EW–luminosity trajectory over multiple variability cycles.
}
\label{f_baldwin}
\end{figure*}
%+++++++++++++++++++++++++++++++++++++++++++++++++++++++++++++++++++++

Throughout this section, we distinguish between the ensemble Baldwin effect (eBeff) and the intrinsic Baldwin effect (iBeff). The eBeff refers to the population-level anti-correlation between emission-line equivalent width (EW) and continuum luminosity measured from single-epoch quasar samples, whereas the iBeff describes the time-resolved EW response to continuum variability within individual sources. The eBeff should therefore be understood as an ensemble projection of intrinsically time-dependent EW–luminosity evolution, rather than a unique relation that must be followed by any individual source at all times.

Figure~\ref{f_baldwin} presents the Baldwin effect for the \civ~emission line in our sample. The left panel shows the rest-frame EW of \civ~as a function of bolometric luminosity for sources observed at two spectroscopic epochs, corresponding to continuum-bright and continuum-dim states. Grey contours indicate the eBeff measured from the SDSS DR16Q quasar sample \citep{Wu2022}. For each source, arrows connect the two epochs, illustrating the two-epoch intrinsic Baldwin response evaluated against the ensemble relation. While a substantial of sources follow the expected eBeff-like behavior, a non-negligible fraction exhibits an anti-eBeff response, in which the EW change between epochs is opposite to the ensemble trend.

The right panel of Figure~\ref{f_baldwin} provides a zoomed-in view of the intrinsic Baldwin evolution (iBeff) for the illustrative source J142308.03+522815.6, based on 73 spectroscopic epochs. An analogous two-epoch intrinsic Baldwin response and an illustrative iBeff example for the \ciii~emission line, corresponding to the same source shown in the right panel of Figure~\ref{f_baldwin}, are presented in Appendix~\ref{sec_beff_c3} (Figures~\ref{f_baldwin_c3} and \ref{f_ibeff_info}).
In this case, the EW–luminosity trajectory reveals a coherent, time-resolved evolution characterized by a zigzag-like pattern in the EW–$L$ plane. Individual brightening and dimming episodes generally proceed along directions broadly consistent with the ensemble Baldwin relation, but with varying local slopes. Notably, successive variability cycles exhibit systematically changing responsivities, with the EW response to continuum variations becoming progressively weaker over time. This behavior naturally produces a composite trajectory in which short-term motions follow eBeff-like directions, while the long-term evolution manifests as a gradual drift across the EW–luminosity plane.

The apparent anti-eBeff responses observed in the two-epoch population view can therefore be understood as a consequence of undersampling this intrinsically time-dependent evolution. 
When only two spectroscopic epochs are available, the measured EW response depends sensitively on which segments of the zigzag trajectory are sampled. Pairs of epochs that straddle different phases of the intrinsic evolution may yield anti-eBeff responses, even though the underlying time-resolved behavior remains broadly consistent with eBeff-like trends. Consistent behavior is also observed for \ciii, further supporting that the presence of anti-eBeff responses in two-epoch measurements is a generic consequence of sparse temporal sampling rather than a line-specific effect (see Appendix~\ref{sec_beff_c3}). 
The slightly higher anti-eBeff fraction observed during brightening episodes is consistent with this picture, as phases with stronger EW responsivities and steeper local slopes are more likely to be undersampled by two-epoch measurements that straddle different segments of the zigzag trajectory.

This interpretation is consistent with previous studies of changing-look quasars, which have reported weak, reversed, or absent Baldwin trends when comparing bright and dim states using two-epoch spectroscopy (e.g., \citealt{Chen2026}). Such results are inherently sensitive to sparse temporal sampling and to which phases of the intrinsic variability cycle are probed.  

\rv{Our results extend this picture by explicitly resolving the time-domain EW–luminosity trajectory for an individual source, demonstrating that anti-eBeff responses naturally arise from undersampling a non-linear, time-dependent intrinsic Baldwin evolution, rather than requiring a fundamentally different physical mechanism. This interpretation is also broadly consistent with the findings of \citet{Ren2022}, who reported an intrinsic Baldwin-like behavior in \clagn based on stacked spectra across different brightness states. While their analysis captures the ensemble-averaged response, our time-resolved spectroscopy reveals that the intrinsic EW–luminosity relation within individual sources is non-stationary and evolves across variability cycles, naturally giving rise to apparent anti-eBeff responses in sparsely sampled, two-epoch measurements.}

Finally, to ensure that our continuum and line measurements entering the Baldwin analysis are on a consistent scale, we compare our measurements at the SDSS epoch with those reported by \cite{Wu2022}. The results of this comparison are presented in Appendix~\ref{sec_fit_global} (Figure~\ref{f_qiaoya}), demonstrating that systematic differences in spectral fitting do not drive the observed Baldwin-related trends. \rvt{We note that the apparent anti-eBeff behavior may be influenced by measurement uncertainties, particularly in the low-EW regime where relative errors are larger. While such uncertainties can introduce additional scatter in the EW–luminosity plane, our conservative quality cuts ($\mathrm{S/N} > 5$ and $\chi^2 < 2.5$) ensure robust line measurements. Moreover, the coherent, time-resolved EW–luminosity trajectory observed in the multi-epoch case suggests that the anti-eBeff behavior is unlikely to be primarily driven by measurement uncertainties, but instead reflects intrinsic variability combined with sparse temporal sampling.}

% -----------------------------------------------------------------------------------
\subsection{Eddington Ratio}\label{sec_edd}

%+++++++++++++++++++++++++++++++++++++++++++++++++++++++++++++++++++++
\begin{figure*}[htbp]
\centering
\includegraphics[width=0.48\linewidth]{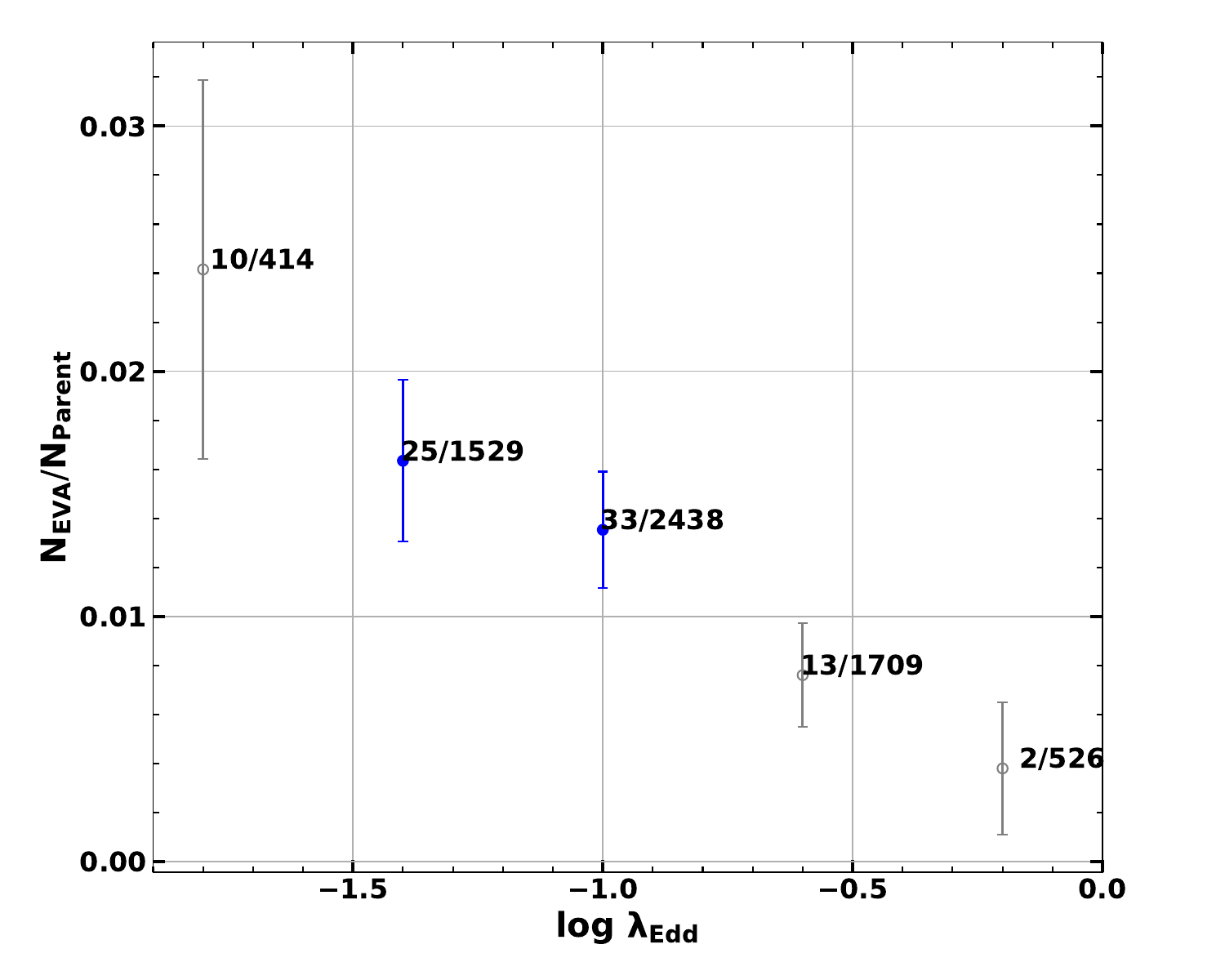}
\includegraphics[width=0.48\linewidth]{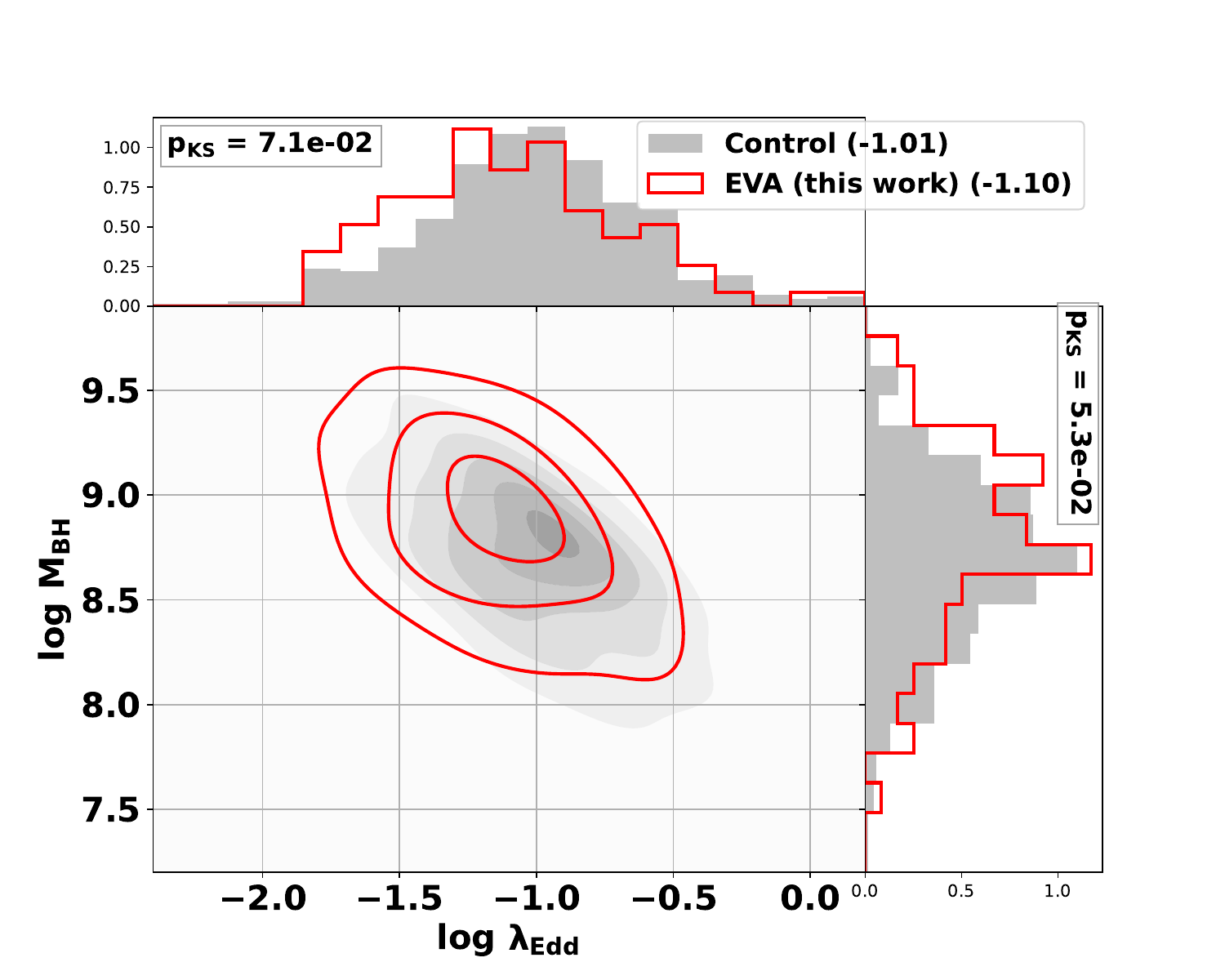}
\includegraphics[width=0.48\linewidth]{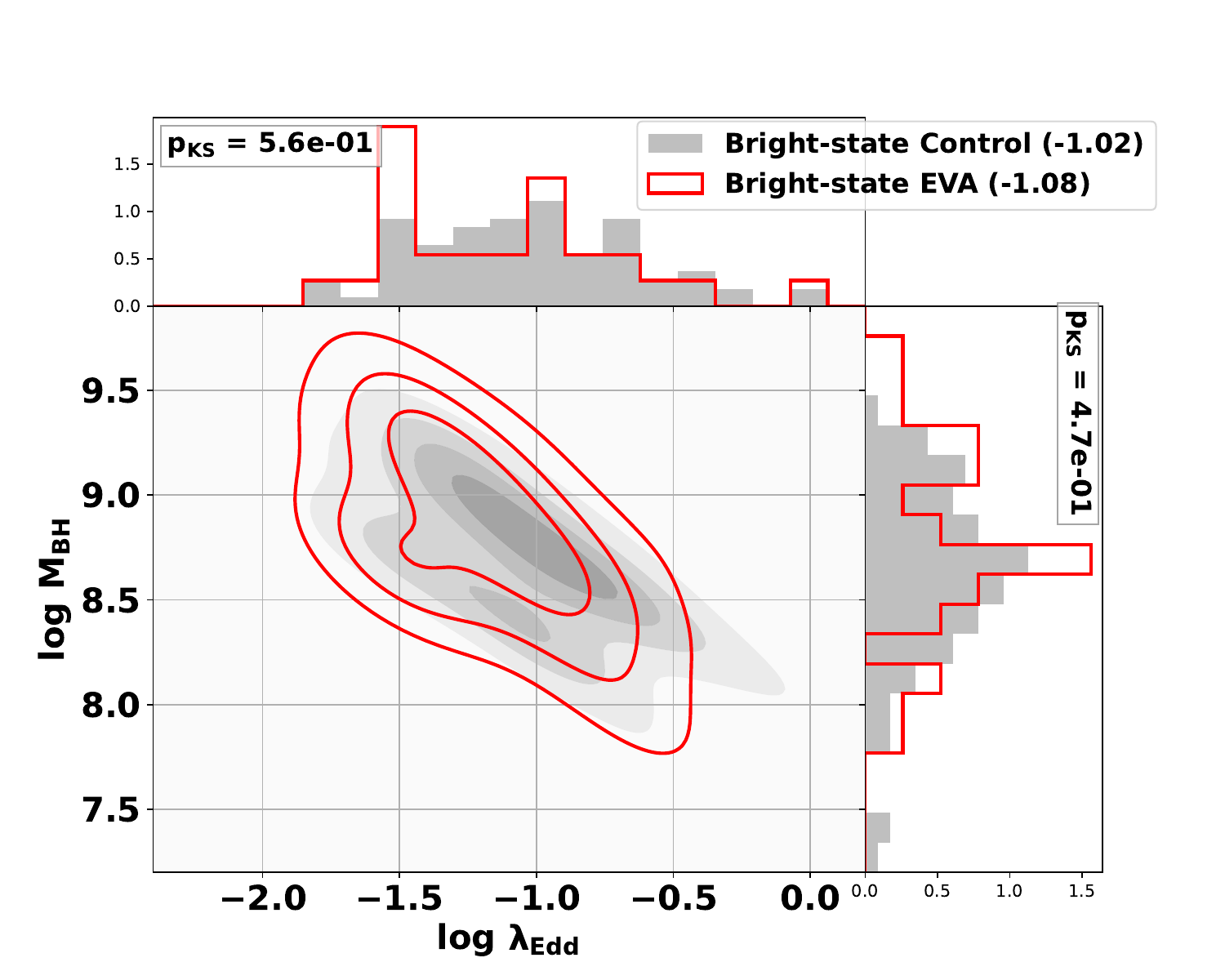}
\includegraphics[width=0.48\linewidth]{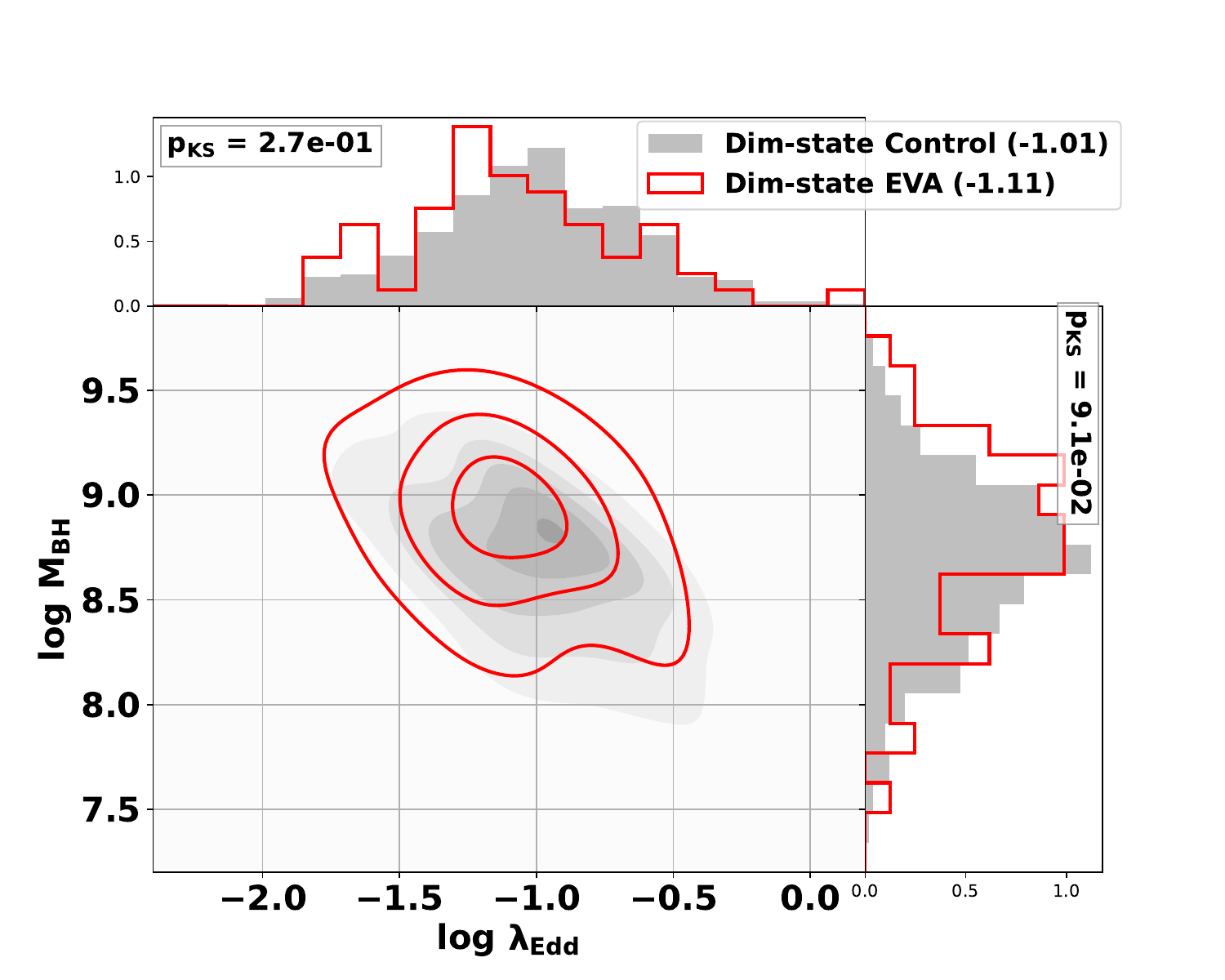}
\caption{
Upper left: The \clagn fraction relative to our full external-spectroscopy $\times$ HETDEX parent sample, shown as a function of the Eddington ratio. Bins with more than 25 \clagn are shown in blue, while bins with lower statistics are shown in gray.
The remaining three panels compare the Eddington ratio and black hole mass between the \clagn sample and their corresponding control sample matched in bolometric luminosity ($L_{\rm bol}$) and redshift.
The upper right panel presents the comparison for the full \clagn sample identified in this work.
The bottom left and bottom right panels show the same comparison for \clagn in the bright and dim states, respectively.
In each of the three 2D comparison panels, the overlaid histograms represent the marginalized distributions of the Eddington ratio and black hole mass.
The median logarithmic Eddington ratio for each sample is indicated in parentheses in the legends.
The p-values from the Kolmogorov--Smirnov tests comparing the \clagn and control distributions are also annotated.
}
\label{f_edd}
\end{figure*}
%+++++++++++++++++++++++++++++++++++++++++++++++++++++++++++++++++++++

Figure~\ref{f_edd} examines the Eddington-ratio properties of the \clagn population.
The upper left panel shows the \clagn fraction as a function of $\lambda_{\rm Edd}$.
Although only a limited number of bins satisfy the threshold for sufficient statistics, the data suggest a weak tendency for higher \clagn fractions at lower Eddington ratios.
Given the measurement uncertainties and limited dynamic range, this trend is not statistically significant, but it motivates a more detailed comparison between the \clagn and control samples.

The remaining three panels compare the distributions of Eddington ratio and black hole mass between the \clagn sample and control samples matched in bolometric luminosity and redshift.
In the upper right panel, the full \clagn population broadly overlaps with its control sample in the $\log \lambda_{\rm Edd}$--$\log M_{\rm BH}$ plane, but shows a mild shift toward lower Eddington ratios.
A Kolmogorov--Smirnov test yields a marginal difference ($p_{\rm KS} \simeq 0.07$), indicating a suggestive but not statistically significant preference for lower $\lambda_{\rm Edd}$ among \clagn.

The bottom panels further separate the comparison into bright- and dim-state \clagno.
Dim-state \clagn exhibit a modest offset toward lower Eddington ratios relative to their matched control sample, with median $\log \lambda_{\rm Edd}$ values differing by $\sim$ 0.1~dex, although the difference is not statistically significant given the limited sample size.
In contrast, bright-state \clagn are statistically indistinguishable from their control counterparts.
These results suggest that any overall tendency toward lower Eddington ratios in the full \clagn sample may be associated with sources observed in their dim states, rather than with enhanced accretion during bright phases.

\rv{All bolometric luminosities, black hole masses, and Eddington ratios used in this section are adopted from \citet{Wu2022}, based on measurements from SDSS spectra, and are therefore available only for sources with SDSS spectroscopic epochs.
This choice is motivated by the limited wavelength coverage of HETDEX spectra (3500--5500~\AA), which often does not fully cover the broad emission lines required for reliable single-epoch estimates of $M_{\rm BH}$ and $\lambda_{\rm Edd}$.
Because the SDSS observations typically precede the HETDEX epochs, brightening \clagn are characterized using their dim-state (SDSS) measurements, while dimming \clagn are characterized using their bright-state (SDSS) measurements.
This approach ensures a homogeneous and physically consistent set of accretion parameters across the full \clagn sample.}

Taken together, the four panels in Figure~\ref{f_edd} indicate that \clagn do not exhibit a strong or statistically significant preference for extreme Eddington ratios.
Instead, they show a mild tendency toward lower $\lambda_{\rm Edd}$ values, broadly consistent with trends reported in previous studies (e.g., \citealt{Zeltyn2024}).
The lack of a significant difference likely reflects a combination of factors, including our operational definition of \clagn based on emission-line variability rather than extreme type transitions, the limited sample size and asymmetric epoch coverage imposed by the use of SDSS spectra, and the substantial uncertainties in single-epoch $M_{\rm BH}$ and $\lambda_{\rm Edd}$ estimates for $z \sim 1.5$ AGN.

% ====================================================================================
\section{Summary} \label{sec_summary}

We identify a large and homogeneous sample of \clagn by combining untargeted and deep HETDEX spectroscopy with multi-epoch spectra from SDSS, DESI, and LAMOST.
The parent sample is constructed by anchoring all candidates to a HETDEX spectroscopic epoch and pairing it with available spectra from SDSS, DESI, and LAMOST for sources identified as AGN in either the HETDEX or the external epoch(s), using a symmetric forward and reverse strategy that does not assume which epoch is brighter or fainter.
A multi-channel selection based primarily on emission-line flux ratios, supplemented by S/N-based criteria and literature-confirmed cases, is applied, followed by extensive visual inspections using contemporaneous photometric light curves.
This procedure yields a final sample of \nclagn \clagn, of which \nclagnnew are newly identified in this work.
Owing to the combination of untargeted and deep HETDEX spectroscopy, our sample extends to systematically higher redshifts than most previous studies of emission-line variable AGN, providing the first statistical census of such variability over $0.5 \lesssim z \lesssim 2.5$, with a redshift distribution peaking near cosmic noon.

% ===================== Updated Summary points (replace items 1--6) =====================
1. %(\S\,3.1; Duty Cycle)
Within this sample, the incidence fraction of \clagn is $f_{\rm \clagn}\sim0.9\%$ relative to the external-spectroscopy $\times$ HETDEX parent AGN sample.
Interpreted operationally, this fraction provides a lower limit on the duty cycle of significant emission-line variability, corresponding to a cumulative timescale of $t_{\rm net,\clagn}\gtrsim10^{4}$--$10^{6}$ yr over a typical quasar lifetime.

2. %(\S\,3.2; Time-scale)
The rest-frame intervals between spectroscopic epochs span $\sim$1--10 yr and constrain the episodic variability timescale.
Brightening ($\Delta T\sim2.2$ yr) and dimming ($\Delta T\sim2.6$ yr) events show statistically indistinguishable distributions, indicating no significant asymmetry in the characteristic variability timescale.

3. %(\S\,3.3; Line Variation)
Emission-line variability is detected across multiple broad lines, including \lya, \civ, \ciii, \mgii, and \hb, when moderate but statistically significant changes are considered.
Most \clagn exhibit coherent variations in line flux or profile shape rather than complete appearance or disappearance of broad lines, reflecting the adopted variability-based operational definition.

4. %(\S\,3.4; Brightening-to-dimming Ratio)
The observed brightening-to-dimming ratio is largely governed by selection effects tied to the depth and availability of external spectroscopy.
At bright magnitudes ($g\lesssim21$), where such effects are minimized, the brightening and dimming fractions are approximately equal, consistent with the symmetric timescale distributions.

5. %(\S\,3.5; Baldwin Effect)
A key result of this work is the characterization of the Baldwin effect in the time domain.
While many \clagn follow the ensemble Baldwin effect (eBeff) between two epochs, a substantial fraction exhibit apparent anti-eBeff responses.
Time-resolved spectroscopy of an individual source reveals that the intrinsic EW--luminosity relation is non-stationary, with the line-to-continuum responsivity systematically evolving from stronger to weaker across successive variability cycles.
Two-epoch measurements therefore probe different segments of this evolving intrinsic Baldwin evolution (iBeff), naturally producing both eBeff-like and anti-eBeff responses through undersampling, without invoking distinct physical mechanisms.

6. %(\S\,3.6; Eddington Ratio)
The \clagn population shows no strong preference for extreme Eddington ratios.
Relative to control samples matched in bolometric luminosity and redshift, \clagn exhibit a mild tendency toward lower $\lambda_{\rm Edd}$ values, driven primarily by sources observed in their dim states, while bright-state \clagn are statistically consistent with the control population.

In summary, emission-line variability in AGN at cosmic noon is best understood as recurrent, episodic accretion variability rather than rare, discrete state transitions.
The symmetry of brightening and dimming timescales and the near-unity brightening-to-dimming ratio at bright magnitudes point to a largely stochastic accretion process.
Crucially, by resolving the intrinsic Baldwin evolution in the time domain, this work demonstrates that apparent anti-eBeff responses arise naturally from undersampling an evolving, non-stationary line responsivity.
These results provide a unified framework for interpreting emission-line variability phenomenology at the peak epoch of cosmic black hole growth.

%%%%%%%%%%%%%%%%%%%%%%%%%%%%%%%%%%%%%%%%%%%%%%%%%%%%%%%%%%%%%
\clearpage
\begin{acknowledgments}
\vspace{0.2in}
\noindent {\bf Acknowledgments:}

We thank the referee for a constructive and insightful report that significantly improved the quality and clarity of this manuscript.

HETDEX is led by the University of Texas at Austin McDonald Observatory and Department of Astronomy with participation from the Ludwig-Maximilians-Universit\"at M\"unchen, Max-Planck-Institut f\"ur Extraterrestrische Physik (MPE), Leibniz-Institut f\"ur Astrophysik Potsdam (AIP), Texas A\&M University, The Pennsylvania State University, Institut f\"ur Astrophysik G\"ottingen, The University of Oxford, Max-Planck-Institut f\"ur Astrophysik (MPA), The University of Tokyo, and Missouri University of Science and Technology. In addition to Institutional support, HETDEX is funded by the National Science Foundation (grant AST-0926815), the State of Texas, the US Air Force (AFRL FA9451-04-2-0355), and generous support from private individuals and foundations.

The Hobby-Eberly Telescope (HET) is a joint project of the University of Texas at Austin, the Pennsylvania State University, Ludwig-Maximilians-Universit\"at M\"unchen, and Georg-August-Universit\"at G\"ottingen. The HET is named in honor of its principal benefactors, William P. Hobby and Robert E. Eberly.

The authors acknowledge the Texas Advanced Computing Center (TACC, \url{http://www.tacc.utexas.edu}) at The University of Texas at Austin for providing high performance computing, visualization, and storage resources that have contributed to the research results reported within this paper. 

Chenxu Liu acknowledges support from the ``Science \& Technology Champion Project'' (202005AB160002) and from two ``Team Projects'' -- the ``Innovation Team'' (202105AE160021) and the ``Top Team'' (202305AT350002), all funded by the ``Yunnan Revitalization Talent Support Program''. This work is also supported by the National Key Research and Development Program of China (2024YFA1611603) and the ``Yunnan Provincial Key Laboratory of Survey Science'' with project No. 202449CE340002.

Xu Liang acknowledges financial support from the Young Talent Project of Yunnan Province and the Yunnan Province Foundation (202401AT070138)
\end{acknowledgments}

% ====================================================================================
\appendix

\section{Compilation of Previously Reported Changing-Look and Extremely Variable AGN}\label{sec_clagn_lit}

%+++++++++++++++++++++++++++++++++++++++++++++++++++++++++++++++++++++
\begin{deluxetable}{cccc}
\tablecaption{Compilation of \nlit AGN reported in the literature to exhibit dramatic spectral variability, including changing-look AGN and extremely variable quasars, compiled from 39 references and ordered by right ascension.\label{t_lit}}
\tablehead{\colhead{name} & \colhead{redshift} & \colhead{transition} & \colhead{refs}}
\startdata
J000048.17+013313.6 & 0.664 & ? & GuoWJ2025a \\
J000116.00+141123.0 & 0.404 & turn-on & MacLeod2019 \\
J000253.52+210109.9 & 0.146 & turn-off & DongQ2024 \\
J000649.36+115450.0 & 0.472 & ? & GuoWJ2025a \\
J000719.90+253128.6 & 0.559 & ? & ZeltynG2024 \\
\enddata
\tablecomments{
Only the first five entries are shown here for guidance. The full table is available in machine-readable form.\\
Column (1): Source name in IAU JHHMMSS.ss$\pm$DDMMSS.s format. 
Column (2): Spectroscopic redshift reported in the literature.  
Column (3): Transition type as reported in the original reference (e.g., ``turn-on'' or ``turn-off''); when no explicit transition classification is provided, this column is set to ``?''. 
Column (4): Literature reference(s) reporting the variability, denoted by short reference codes; multiple references for the same source are separated by semicolons.\\
}
\end{deluxetable}
%+++++++++++++++++++++++++++++++++++++++++++++++++++++++++++++++++++++

This appendix presents a compilation of AGN reported in the literature to exhibit dramatic spectral variability, including CL-AGN and closely related populations such as extremely variable quasars (EVQs). This compilation includes sources reported up to 2025 June, and is intended to serve as a literature reference catalog rather than a redefinition of the CL-AGN/EVQ class.

We emphasize that the classification schemes adopted in the literature are not uniform: different studies apply different operational definitions when identifying CL-AGN or EVQs, depending on whether the emphasis is placed on the appearance/disappearance of broad emission lines, large-amplitude continuum variability, or extreme emission-line flux changes.

In this compilation, we do not attempt to reclassify sources under a unified definition. Instead, all objects are included as reported in the original references, and the transition type is recorded when explicitly specified; otherwise it is left undefined (“?”).

This table is intended to provide a comprehensive literature reference for previously reported AGN with strong spectral variability, rather than to define a homogeneous CL-AGN sample under a single physical criterion.

The compilation presented in Table~\ref{t_lit} includes AGN reported in the literature
to exhibit dramatic spectral variability, encompassing both changing-look AGN (CL-AGN)
and closely related populations such as extremely variable quasars (EVQs).
We do not attempt to reclassify sources under a unified definition; instead,
all objects are included as reported by the original studies.

The short reference codes used in Table~\ref{t_lit} correspond to the following publications:
AiYL2020 (\citealt{AiYL2020}),
CharltonP2019 (\citealt{CharltonP2019}),
DongQ2024 (\citealt{DongQ2024}),
FrederickS2019 (\citealt{FrederickS2019}),
GrahamM2020 (\citealt{Graham2020}),
GreenP2022 (\citealt{Green2022}),
GuoHX2019 (\citealt{GuoH2019}),
GuoWJ2024 (\citealt{GuoWJ2024}),
GuoWJ2025a (\citealt{GuoWJ2025b}),
GuoWJ2025b (\citealt{GuoWJ2025a}),
HonW2020 (\citealt{Hon2020}),
JinJJ2022 (\citealt{Jin2022}),
LamassaS2015 (\citealt{LaMassa2015}),
LiJ2023 (\citealt{LiJ2023}),
LiuWJ2021 (\citealt{LiuWJ2021}),
Lopez-Navas2023 (\citealt{Lopez-Navas2023}),
LuKX2025 (\citealt{Lu2025}),
MacLeod2016 (\citealt{MacLeod2016}),
MacLeod2019 (\citealt{MacLeod2019}),
MarinF2019 (\citealt{Marin2019}),
ParkerML2019 (\citealt{Parker2019}),
RuanJ2016 (\citealt{RuanJ2016}),
RuncoJ2016 (\citealt{Runco2016}),
RunnoeJ2016 (\citealt{Runnoe2016}),
ShengZF2017 (\citealt{Sheng2017}),
ShengZF2020 (\citealt{ShengZF2020}),
WangJ2018 (\citealt{WangJ2018}),
WangJ2019 (\citealt{WangJ2019}),
WangJ2020b (\citealt{WangJ2020}),
WangJ2022 (\citealt{WangJ2022}),
WangJ2023 (\citealt{WangJ2023}),
WangS2024 (\citealt{WangS2024}),
WangS2025 (\citealt{WangS2025}),
YangQ2018 (\citealt{YangQ2018}),
YangQ2025 (\citealt{YangQ2025}),
YuX2020 (\citealt{YuX2020}),
ZeltynG2022 (\citealt{Zeltyn2022}),
ZeltynG2024 (\citealt{Zeltyn2024}),
and ZhuLT2024 (\citealt{ZhuLT2024}). 
%Note that the short reference codes (e.g., GuoWJ2025a/b) are internal labels used for convenience in this table and do not necessarily correspond to the alphabetical ordering of the BibTeX ``a/b'' suffixes in the citation list.

% ====================================================================================
\section{Validation of Spectral Measurements and Line Fitting}\label{sec_fit}

%In this appendix, we present representative examples to illustrate the behavior of our spectral fitting procedure, particularly in low–signal-to-noise regimes relevant to the line-ratio–based selection described in Section~\ref{sec_identification}.

In this appendix, we provide validation tests for the spectral measurements and line-fitting procedures adopted in this work, which form the foundation of our emission-line variability (EVA) selection. 
The goals of this appendix are twofold: (1) to demonstrate the consistency of our measurements with previous literature results based on independent fitting pipelines, and (2) to illustrate the robustness and conservative behavior of our fitting strategy in both high– and low–signal-to-noise regimes relevant to the EVA selection.

In Section~B.1, we compare our continuum and emission-line measurements against those reported by \cite{Wu2022} for the same SDSS spectral epochs, verifying that our measurements are free of systematic offsets over the full dynamic range. 
In Section~B.2, we present a representative example to illustrate how the fitting procedure reliably reproduces complex, blended emission-line profiles when genuine signals are present, while intentionally suppressing spurious line detections in noise-dominated regions. 
Together, these tests demonstrate that the spectral measurements and fitting behavior underlying the \clagn selection are both consistent with prior work and deliberately conservative in the presence of low signal-to-noise data.

\subsection{Consistency with Literature Measurements}\label{sec_fit_global}

\begin{figure*}[htbp]
\centering
\includegraphics[width=0.49\linewidth]{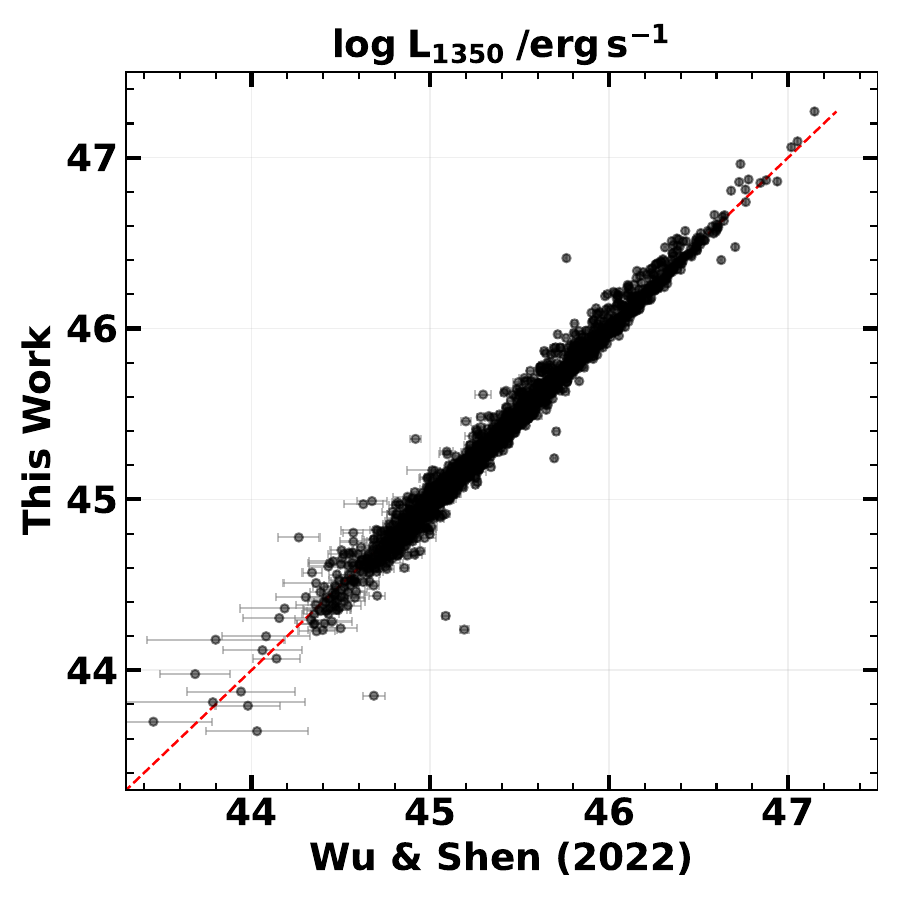}
\includegraphics[width=0.49\linewidth]{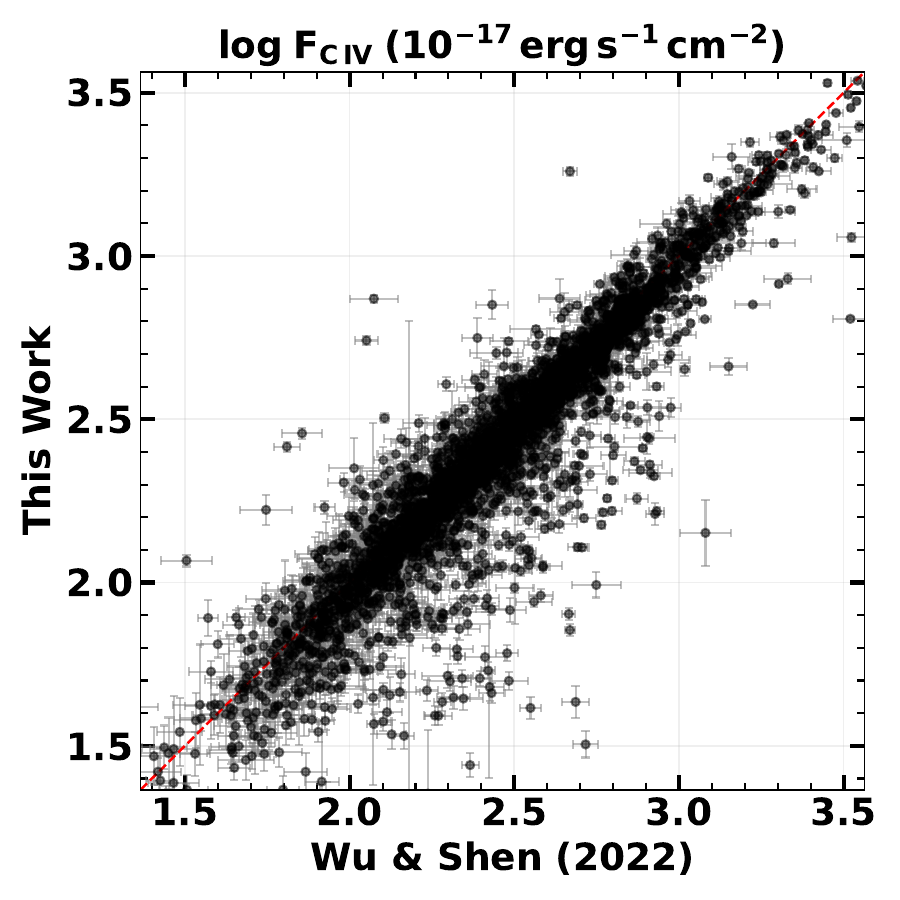}
\caption{Comparison of the continuum luminosity at 1350\,\AA\ (left) and the \civ~line flux (right) measured in this work with those reported by \cite{Wu2022}, evaluated at the same SDSS spectral epoch for each source. These two quantities are the key parameters used in our Baldwin-effect analysis (Figure~\ref{f_baldwin}). The red dashed line indicates the one-to-one relation. The close agreement and absence of systematic offsets demonstrate that our spectral measurements are consistent with previous work.
}
\label{f_qiaoya}
\end{figure*}

We first verify that our spectral measurements are consistent with previous studies. Figure \ref{f_qiaoya} compares the continuum luminosity at 1350 Å and the \civ~line flux measured in this work with those reported by \cite{Wu2022}, showing excellent agreement over the full dynamic range.

%+++++++++++++++++++++++++++++++++++++++++++++++++++++++++++++++++++++
%+++++++++++++++++++++++++++++++++++++++++++++++++++++++++++++++++++++

\subsection{Fitting Robustness and Low–Signal-to-Noise Treatment}\label{sec_fit_individual}

\begin{figure*}[htbp]
\centering
\includegraphics[width=0.95\linewidth]{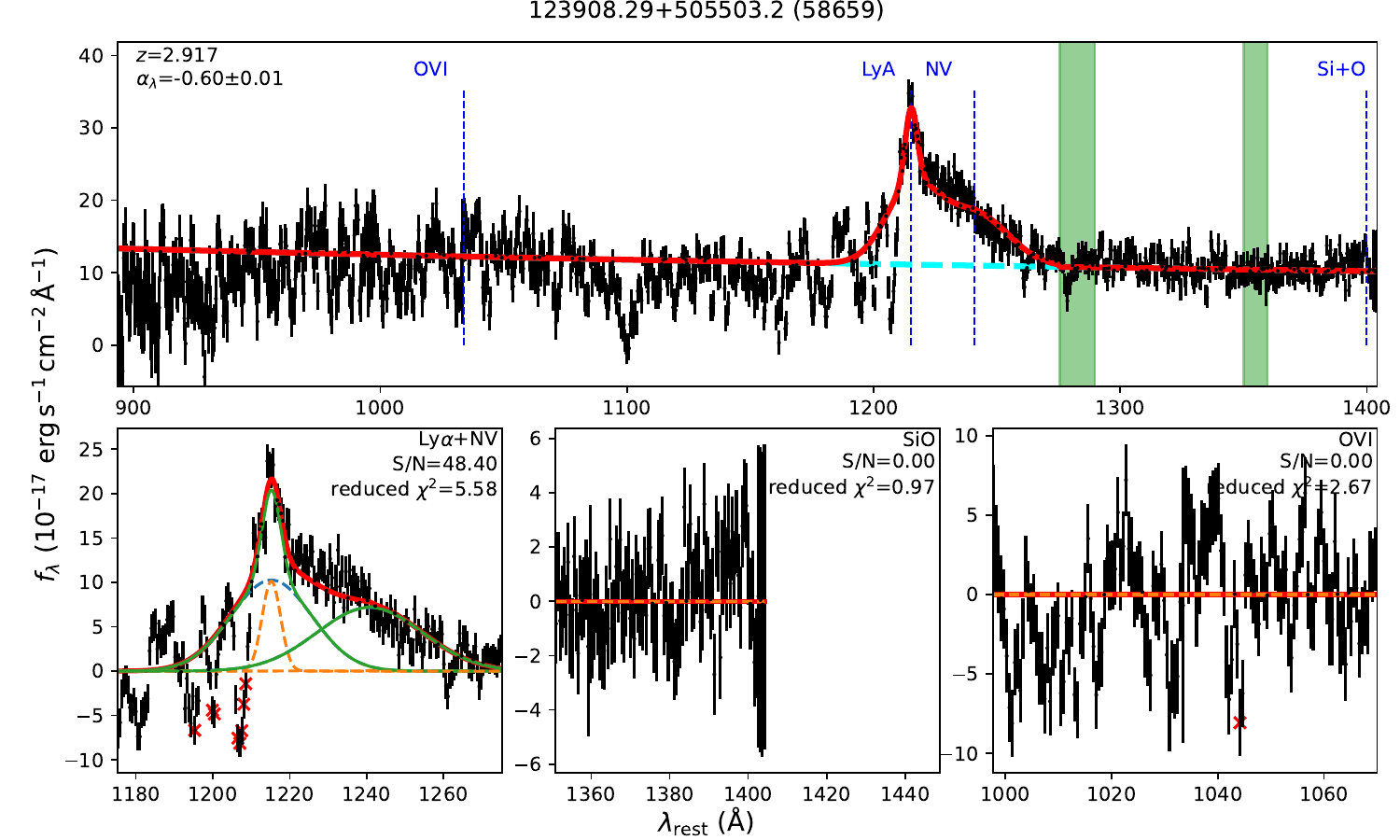}
\caption{Representative example illustrating the behavior of the spectral fitting procedure in both high- and low–signal-to-noise regimes. The spectrum shown corresponds to the HETDEX observation of J123908.29+505503.2 at MJD = 58659 and spans the full rest-frame wavelength coverage available from HETDEX.
The top panel shows the rest-frame spectrum together with the best-fitting continuum and emission-line model; the green shaded regions indicate the wavelength windows adopted for continuum fitting.
In the Ly$\alpha$+N,V region (lower left), the complex line profile is well reproduced by the multi-component fit, yielding a high signal-to-noise detection.
In contrast, for weak or noise-dominated features such as \sio~and \ovi~(lower middle and right), the fitting naturally degenerates to a linear continuum when ${\rm SNR}_{\rm pix}<1$, and all derived line parameters are set to zero by construction.
This behavior prevents spurious line-flux measurements from noise-dominated regions while preserving robust fits for genuinely detected emission lines.
}
\label{f_specfit}
\end{figure*}

We further illustrate the behavior of the fitting procedure in low–signal-to-noise regimes relevant to the EVA selection. %As shown in Figure~\ref{f_specfit}, when ${\rm SNR}_{\rm pix}<1$, the multi-component line model naturally degenerates to a linear continuum, and all derived line parameters are set to zero by construction.
The example shown in Figure~\ref{f_specfit} demonstrates two key aspects of our spectral modeling strategy. 
First, in regions with genuinely detected emission features, such as the Ly$\alpha$+N\,V complex, the multi-component fitting procedure is able to reproduce the observed line profiles robustly, even in the presence of blending and absorption. 
The fitting methodology for complex ultraviolet emission-line regions follows that developed and validated in our previous HETDEX AGN catalog studies \citep{Liu2022a, Liu2025}.

Second, for spectral regions dominated by noise, where the pixel-level signal-to-noise ratio falls below ${\rm SNR}_{\rm pix}=1$, the line model is designed to degenerate to a linear continuum. In such cases, all line-related parameters are set to zero, ensuring that noise fluctuations do not produce artificial line detections or inflated line-flux ratios. 
This conservative treatment is essential for the robustness of the emission-line variability selection adopted in this work.

%+++++++++++++++++++++++++++++++++++++++++++++++++++++++++++++++++++++
%+++++++++++++++++++++++++++++++++++++++++++++++++++++++++++++++++++++

% ====================================================================================
\section{Examples Excluded During Visual Inspection}\label{sec_vi_rej}

As described in Section~\ref{sec_vi}, all emission-line variability (EVA) candidates were subjected to a final stage of visual inspection to ensure consistency between spectroscopic variability and available photometric constraints, as well as to exclude cases where the observed spectral changes are not physically interpretable within the EVA framework adopted in this work.

\begin{figure*}[htbp]
\centering
\includegraphics[width=0.49\linewidth]{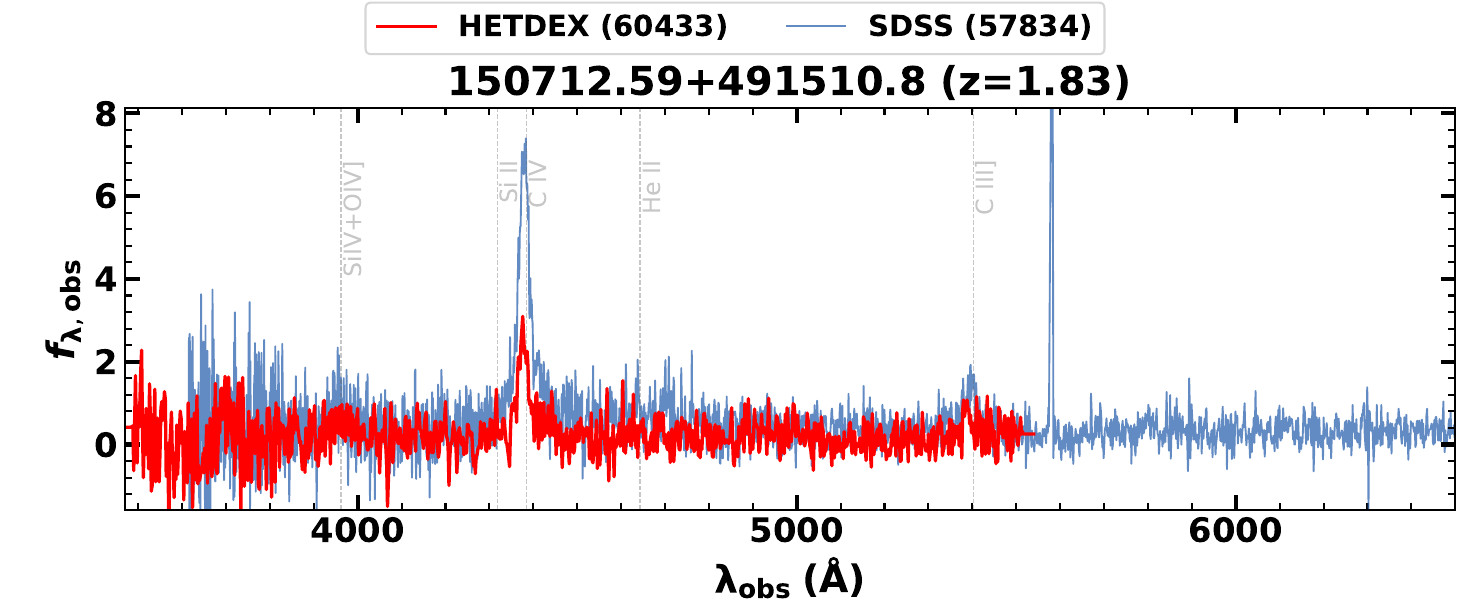}
\includegraphics[width=0.49\linewidth]{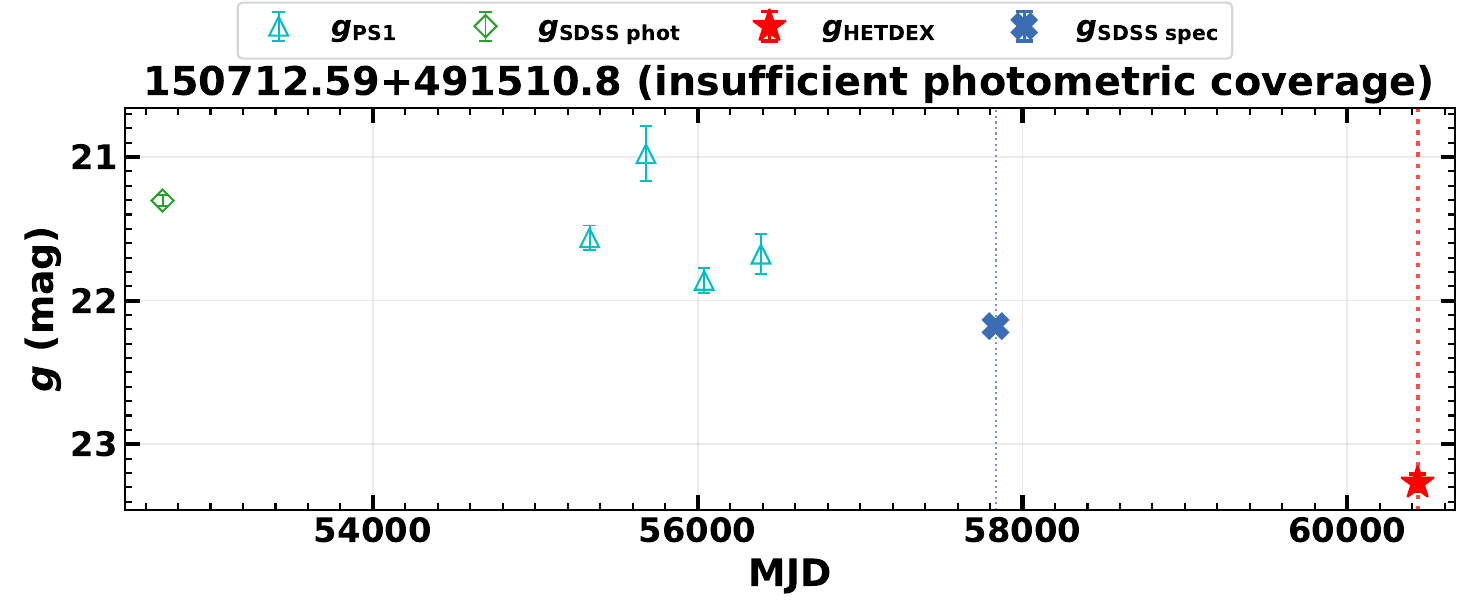}\\
\includegraphics[width=0.49\linewidth]{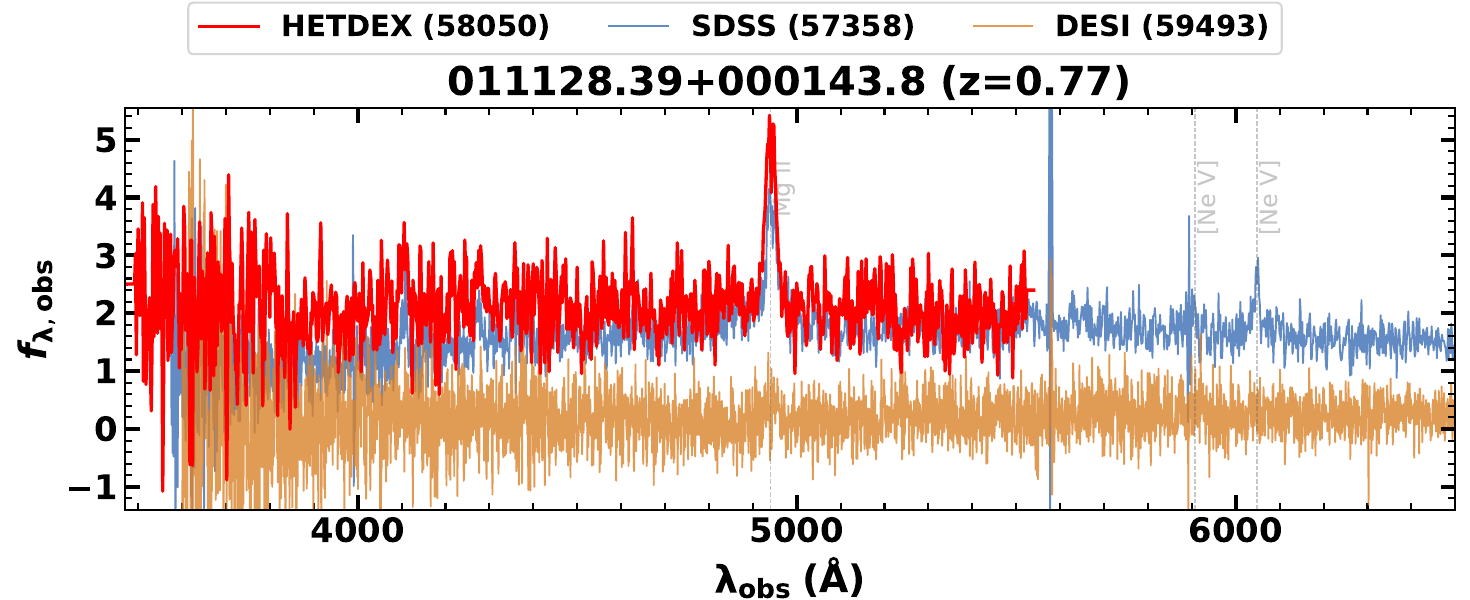}
\includegraphics[width=0.49\linewidth]{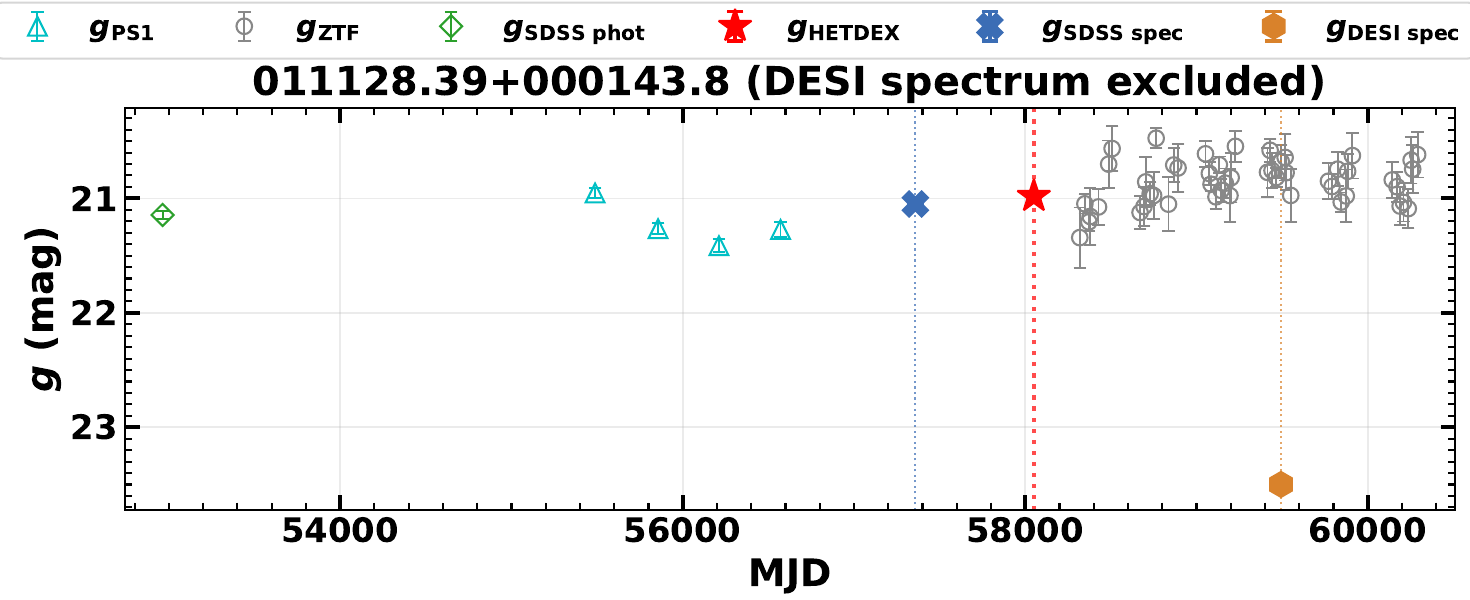}\\
\includegraphics[width=0.49\linewidth]{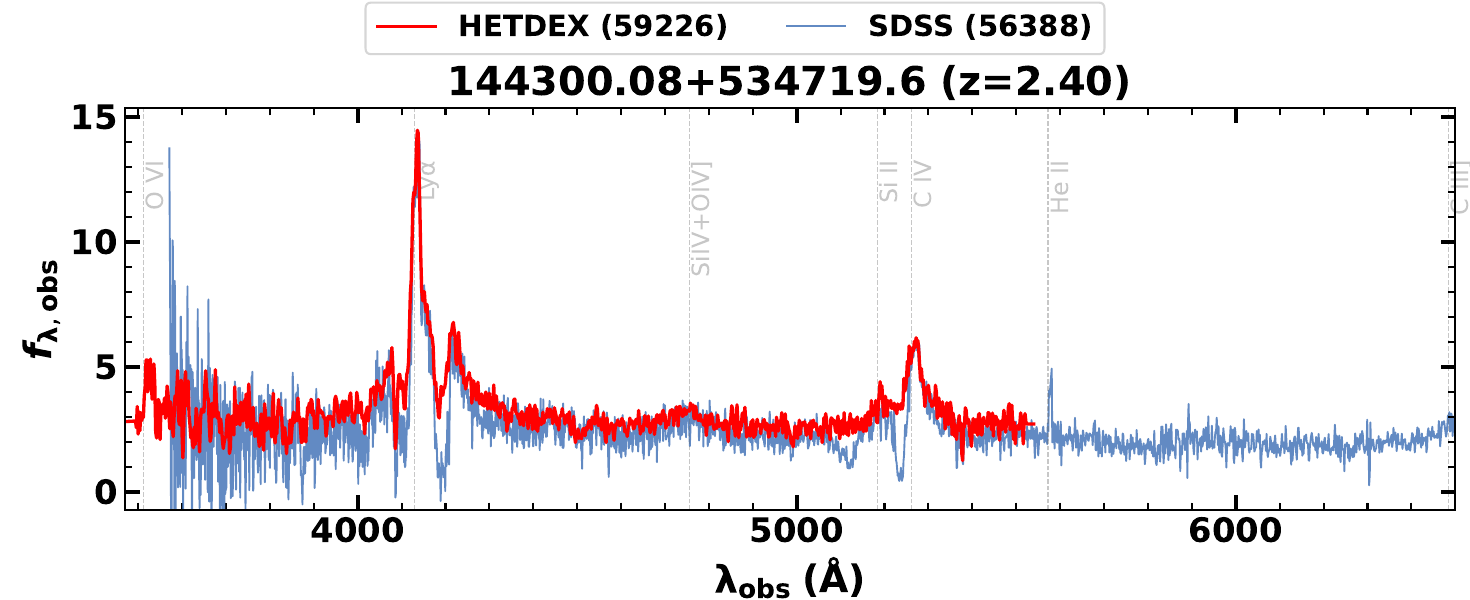}
\includegraphics[width=0.49\linewidth]{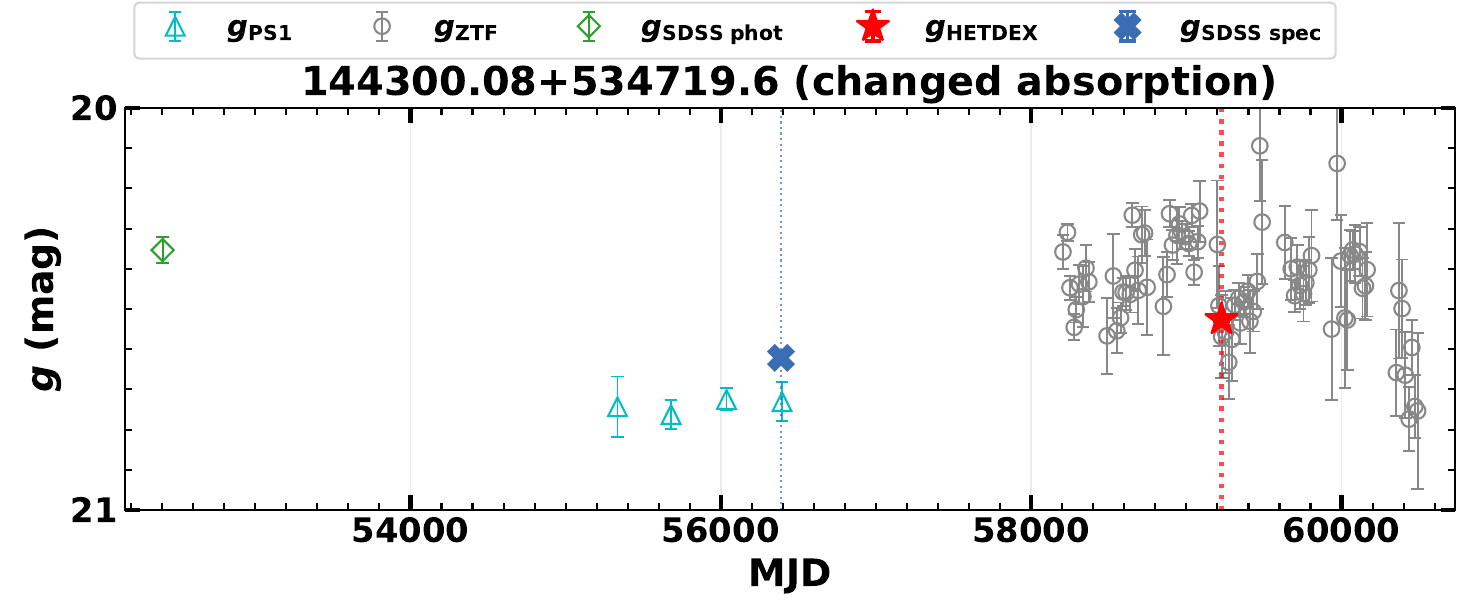}\\
\caption{Representative examples of sources excluded during visual inspection.
The labels indicate the specific reasons for exclusion under the emission-line variability (EVA) selection framework adopted in this work (see Section~\ref{sec_vi}), including insufficient photometric coverage, exclusion of a given spectral epoch, or line flux variability dominated by absorption features.
These exclusions reflect consistency requirements between spectroscopic and photometric behavior, rather than data-quality limitations of the underlying surveys.}
\label{f_vi_rej}
\end{figure*}

Figure~\ref{f_vi_rej} presents representative examples of sources that were removed during this process. 
These cases illustrate common exclusion scenarios, including (i) insufficient photometric coverage to verify the spectroscopic variability, (ii) exclusion of specific spectral epochs due to calibration or extraction issues, and (iii) apparent variability dominated by changes in absorption features rather than emission-line fluxes.

We emphasize that these exclusions do not indicate poor data quality in the original surveys (e.g., SDSS, DESI, or HETDEX), nor do they contradict previous classifications in the literature. 
Instead, they reflect the conservative requirements imposed by our EVA-based selection, which prioritizes reproducible emission-line flux changes that can be robustly linked to physically interpretable nuclear variability.

%+++++++++++++++++++++++++++++++++++++++++++++++++++++++++++++++++++++
%+++++++++++++++++++++++++++++++++++++++++++++++++++++++++++++++++++++

% ====================================================================================

\section{Baldwin Effect of \ciii}\label{sec_beff_c3}
%+++++++++++++++++++++++++++++++++++++++++++++++++++++++++++++++++++++
\begin{figure*}[htbp]
\centering
\includegraphics[width=0.49\linewidth]{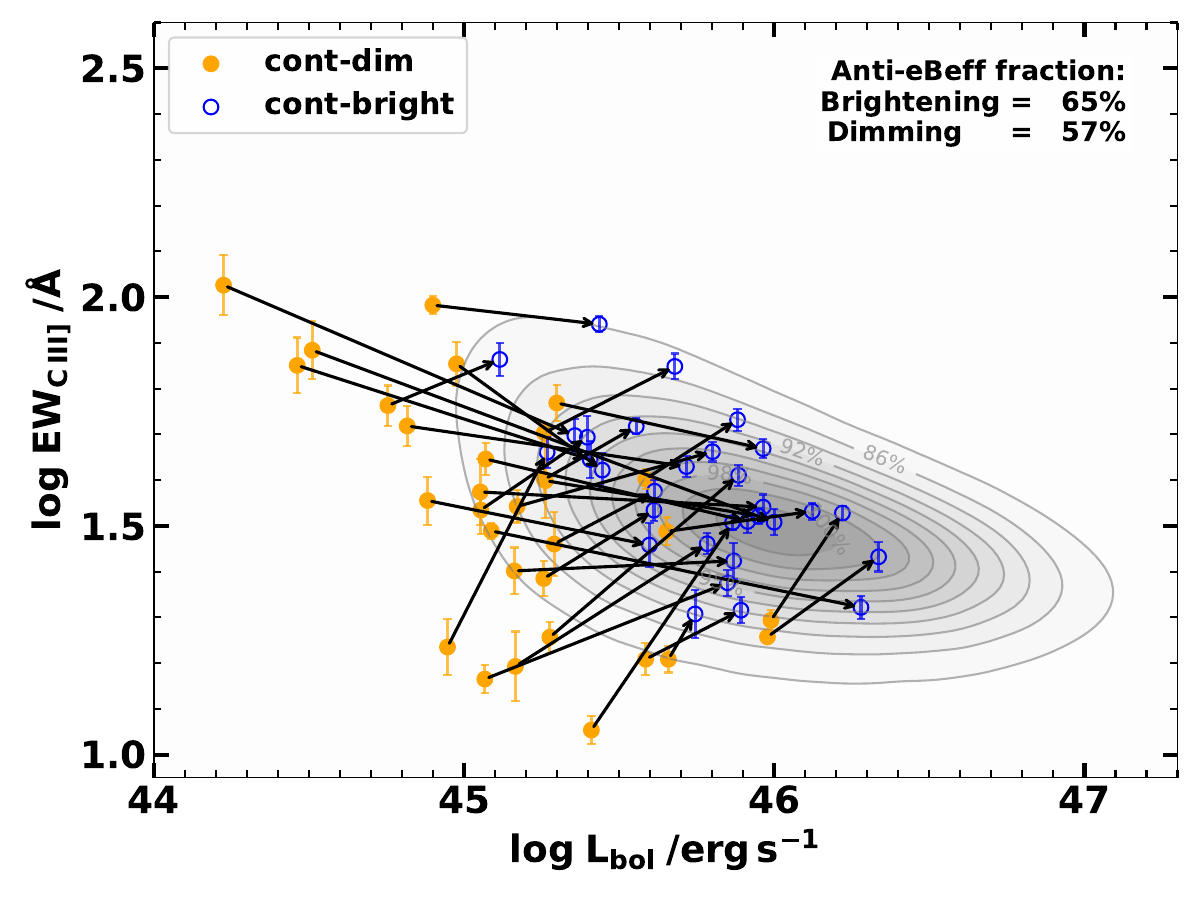}
\includegraphics[width=0.49\linewidth]{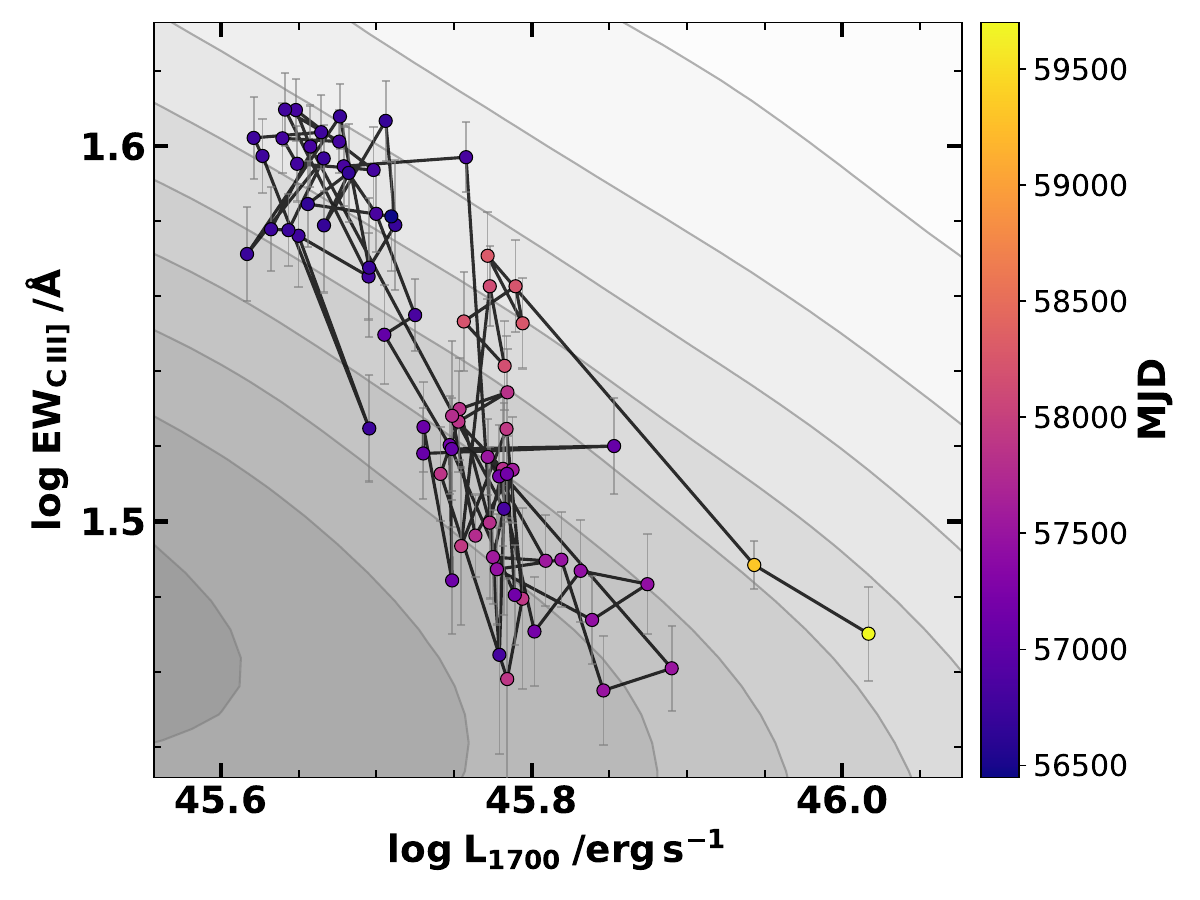}
\caption{Same as Figure~\ref{f_baldwin}, but for the \ciii~emission line, showing the two-epoch intrinsic Baldwin response (evaluated against the ensemble Baldwin relation, eBeff) for the \clagn sample (left) and the intrinsic Baldwin evolution (iBeff) of the illustrative source J142308.03+522815.6 (right).}
\label{f_baldwin_c3}
\end{figure*}
%+++++++++++++++++++++++++++++++++++++++++++++++++++++++++++++++++++++

%+++++++++++++++++++++++++++++++++++++++++++++++++++++++++++++++++++++
\begin{figure*}[htbp]
\centering
\includegraphics[width=0.49\linewidth]{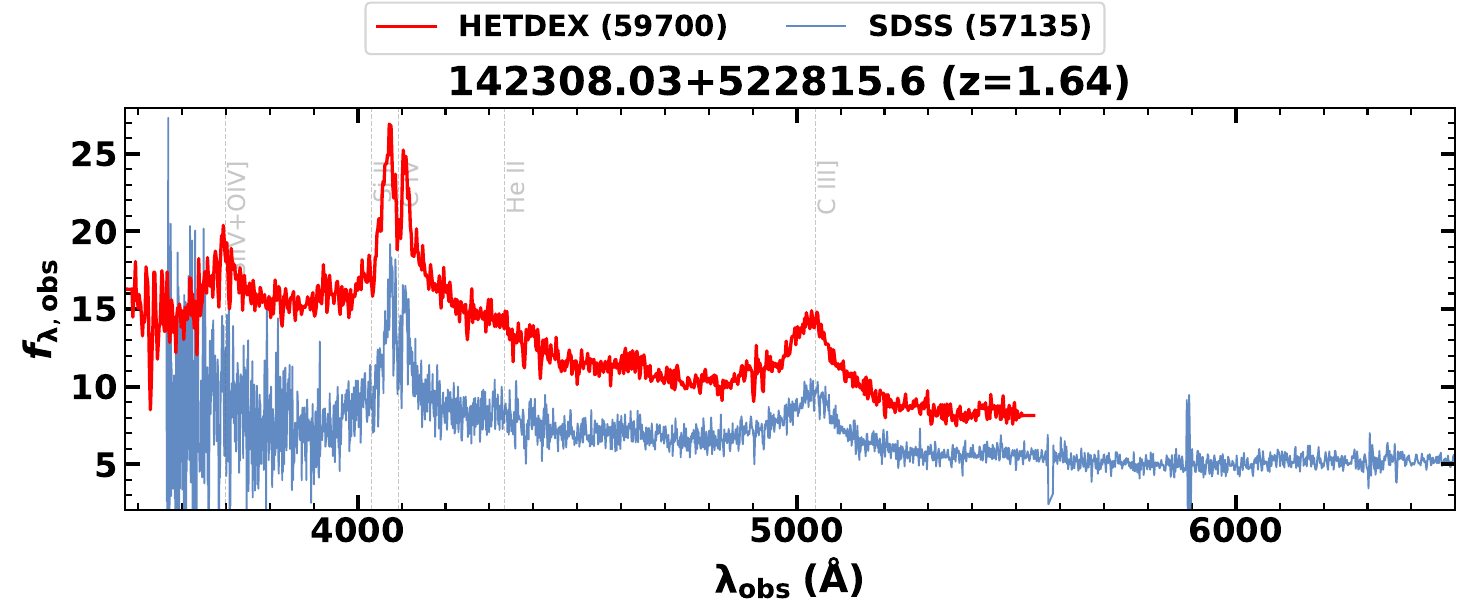}
\includegraphics[width=0.49\linewidth]{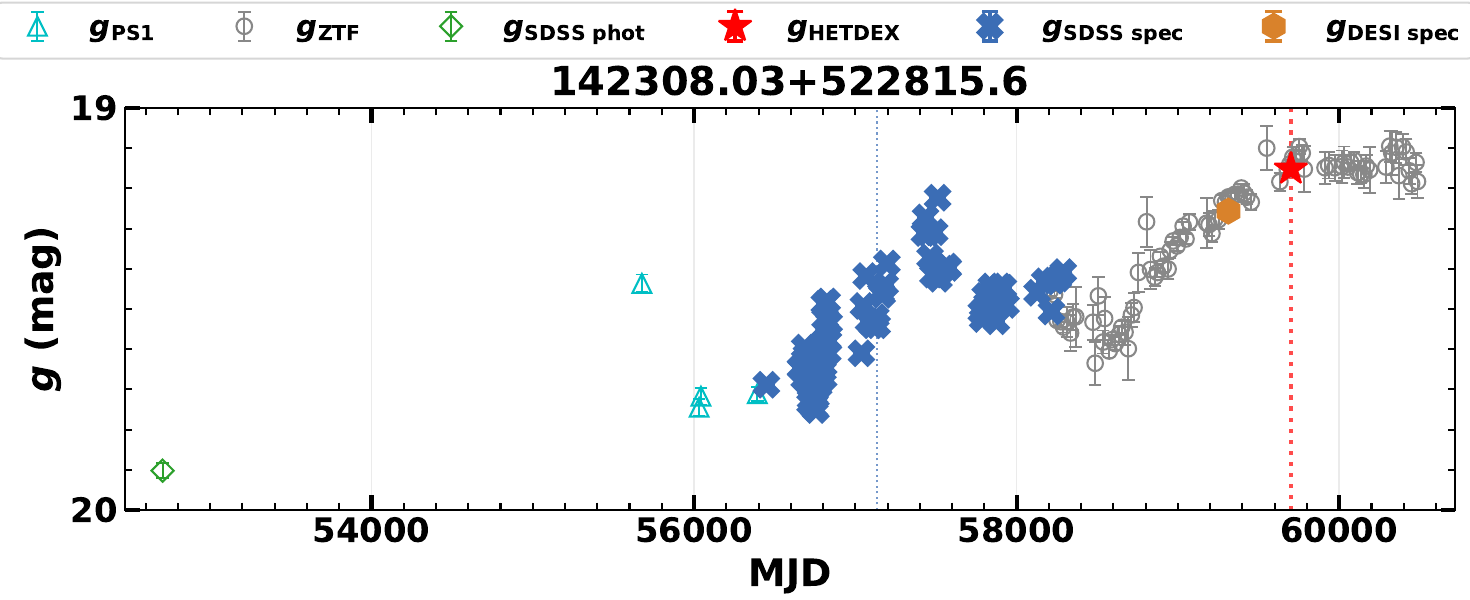}
\caption{Multi-epoch spectroscopy and photometric variability of the iBeff example source J142308.03+522815.6. Left panels: Comparison between representative HETDEX spectra (red) and an external SDSS spectrum (blue), shown in the observed-frame wavelength.
Right panels: Multi-epoch $g$-band light curves compiled from SDSS imaging, PS1, and ZTF, together with pseudo-$g$ magnitudes derived from the SDSS and HETDEX spectra.
Vertical dotted lines indicate the epochs of the corresponding spectroscopic observations in the left panel.
This figure provides the time-domain context for the intrinsic Baldwin evolution (iBeff) of this source discussed in Figures~\ref{f_baldwin} and~\ref{f_baldwin_c3}.
}
\label{f_ibeff_info}
\end{figure*}
%+++++++++++++++++++++++++++++++++++++++++++++++++++++++++++++++++++++

In this appendix, we present an analogous analysis of the Baldwin effect for the \ciii\ emission line, complementing the main results for \civ\ discussed in Section~\ref{sec_baldwin}. The purpose of this section is twofold: (i) to examine whether the two-epoch intrinsic Baldwin responses observed for \civ\ are also present in a lower-ionization ultraviolet line, and (ii) to assess whether the interpretation based on sparse sampling of a time-dependent intrinsic Baldwin evolution is line-dependent.

Figure~\ref{f_baldwin_c3} shows the rest-frame EW of \ciii\ as a function of luminosity for the same \clagn\ sample used in Figure~\ref{f_baldwin}. As in the \civ\ case, the left panel presents the two-epoch intrinsic Baldwin responses evaluated against the ensemble Baldwin relation (eBeff), while the right panel displays the time-resolved intrinsic Baldwin evolution (iBeff) of the illustrative source J142308.03+522815.6, which is identical to the example shown in the right panel of Figure~\ref{f_baldwin}. The \ciii\ measurements reveal a qualitatively similar behavior to that observed for \civ, including eBeff-like short-term motions, varying local responsivities, and the presence of apparent anti-eBeff responses in two-epoch projections.

To provide additional context for the time-domain interpretation, Figure~\ref{f_ibeff_info} presents the multi-epoch spectroscopy and photometric light curves of J142308.03+522815.6. The comparison between representative HETDEX and SDSS spectra, together with the long-term $g$-band light curve, illustrates that the intrinsic Baldwin evolution inferred from the EW--luminosity plane is accompanied by coherent continuum variability over multiple cycles. This confirms that the \ciii\ iBeff behavior arises from genuine time-dependent spectral evolution rather than from measurement noise or calibration differences between epochs.

The consistency between the \civ\ and \ciii\ results supports the conclusion that apparent anti-eBeff responses seen in two-epoch measurements are a generic consequence of sparse temporal sampling of a non-linear, time-dependent intrinsic Baldwin evolution, rather than a line-specific effect. This appendix therefore reinforces the physical interpretation presented in Section~\ref{sec_baldwin}.

%%%%%%%%%%%%%%%%%%%%%%%%%%%%%%%%%%%%%%%%%%%%%%%%%%%%%%%%%%%%%
\bibliography{sample631}{}
\bibliographystyle{aasjournal}

\end{document}